\RequirePackage{fix-cm}
\documentclass[onecolumn,epjc3]{svjour3}  
\smartqed
\RequirePackage{graphicx}
\RequirePackage{amsmath}
\RequirePackage{multicol}
\usepackage{setspace}
\usepackage{xcolor}
\usepackage[numbers,sort&compress]{natbib}
\usepackage{caption}
\usepackage{setspace}
\usepackage{appendix}
\begin{document}

\title{Motions of Test Particles in Gravitational Field, Perturbations and Greybody Factor of Bardeen-like AdS Black Hole with Phantom Global Monopoles
}

\author{
\begin{minipage}[t]{\textwidth}
\centering
\normalsize 
Faizuddin Ahmed\thanksref{e1, addr1}, Ahmad Al-Badawi\thanksref{e2, addr2}, \.{I}zzet Sakall{\i}\thanksref{e3, addr3}, Sara Kanzi\thanksref{e4, addr4}
\end{minipage}
}

\thankstext{e1}{\bf e-mail: faizuddinahmed15@gmail.com}
\thankstext{e2}{\bf e-mail: ahmadbadawi@ahu.edu.jo}
\thankstext{e3}{\bf e-mail: izzet.sakalli@emu.edu.tr (Corresp. author)}
\thankstext{e4}{\bf e-mail: sara.kanzi@final.edu.tr}

\institute{
Department of Physics, University of Science \& Technology Meghalaya, Ri-Bhoi, Meghalaya, 793101, India \label{addr1}
\and
Department of Physics, Al-Hussein Bin Talal University 71111, Ma’an, Jordan  \label{addr2}
\and Physics Department, Eastern Mediterranean University, Famagusta 99628, North Cyprus via Mersin 10, Turkey
\label{addr3}
\and Faculty of Engineering, Final International University, North Cyprus via Mersin 10, Kyrenia 99320, Turkey
\label{addr4}
}

\date{Received: date / Accepted: date}

\titlerunning{Motions and Greybody Factor of Bardeen-like AdS BH}
\authorrunning{Ahmed et al.}
\maketitle

\begin{abstract}
We investigate the dynamics of test particles, perturbations, and greybody factors within the framework of a Bardeen-like AdS black hole (BH) with a phantom global monopole. This study explores the interactions between nonlinear electrodynamics, the energy scale of symmetry breaking, and space-time topology. We analyze the geodesic motion of null and time-like particles, deriving effective potentials that describe their trajectories. Utilizing the Regge-Wheeler potential, we calculate the quasinormal modes (QNMs) for scalar, vector, and tensor perturbations, applying the sixth-order WKB approximation. Our findings highlight how the Bardeen-like parameter ($\mathrm{b}$) and the energy scale of symmetry breaking, characterized by the parameter ($\eta$), influence the QNM spectra, with potential implications for gravitational wave observations. We also examine greybody factors, focusing on the transmission and reflection coefficients for scalar and axial fields, and employ semi-analytic techniques to derive precise bounds. Furthermore, we assess the thermodynamic stability of the BH, emphasizing the role of these parameters in phase transitions and stability criteria.



\end{abstract}

\section{Introduction}
\label{sec1}
The study of BHs within the framework of modified theories of gravity \cite{Capozziello:2011et,Clifton:2011jh,Nojiri:2010wj,Joyce:2014kja,Tsujikawa:2010zza} has remained a cornerstone of contemporary research in theoretical physics. Among various solutions, the Bardeen-like anti-de Sitter (AdS) BH with a phantom global monopole (PGM) \cite{Hamil:2024,Kusuma:2021,Zheng:2024,Hu:2020} has garnered considerable attention due to its ability to resolve certain pathological issues present in classical general relativity (GR). Unlike the Schwarzschild or Kerr solutions, which often contain singularities where spacetime curvature diverges, the Bardeen-like AdS BH represents a class of regular BH solutions. These solutions avoid the central singularity by introducing a modified energy-momentum tensor derived from nonlinear electrodynamics (NED) \cite{Kruglov:2021,Russo:2024,Deser:1969,Kruglov:2023,Kruglov:2024,AA2,AA5,AA6}. The inclusion of a PGM, characterized by a symmetry-breaking energy scale $\eta$, further enriches the spacetime geometry, offering insights into how topological defects and exotic matter influence BH physics \cite{AA7,AA9,AA14}. 

The Bardeen-like AdS BH with a PGM has been extensively analyzed in various contexts. Its geodesic structure reveals intricate behaviors of both time-like and null particles \cite{AA14,AA15}, which are essential for understanding astrophysical phenomena such as gravitational lensing and accretion disk dynamics \cite{Bisnovatyi-Kogan:2022,Boero:2021,Best:2022,Huang:2024,Universe:2024}. Additionally, its thermodynamic properties, including the Hawking temperature, heat capacity, and entropy, provide an effective testing ground for theories of quantum gravity and thermodynamic stability \cite{AA10,AA11,AA12}. The Regge-Wheeler (RW)-potential \cite{Regge:1957,Zerilli:1970,Chandrasekhar:1975,Karlovini:2002,Mukohyama:2022} associated with perturbations in these spacetimes has also been a subject of interest, as it helps characterize QNMs, which are indeed the characteristic oscillations of BHs and other compact objects that arise due to perturbations in the spacetime geometry. These modes, governed by the RW and Zerilli equations for various BHs, exhibit complex frequencies whose real parts correspond to oscillation rates and imaginary parts determine damping due to energy loss via gravitational waves. The study of QNMs plays a crucial role in gravitational wave astronomy \cite{AA16,AA17,Konoplya}, particularly in the ringdown phase of BH mergers observed by LIGO and Virgo \cite{Kokkotas:1999, Berti:2009, Konoplya:2011, Nollert:1999, Vishveshwara:1970, LIGO:2021, Siegel:2023}. Therefore, the presence of a PGM is expected to modify QNMs significantly, leading to observable signatures that could be tested in future experimental setups \cite{KY,ZQD,ELBJ}.

Despite these advancements, there remains a critical need to explore the greybody factors \cite{Fernando:2008,Konoplya:2010,Das:1997,Sakalli:2022,Kanzi:2020,Kanzi:2023,Birmingham:2003,Cardoso:2001,Badawi1,Badawi2} of these BHs, particularly focusing on how the spacetime parameters influence the reflection and transmission coefficients for perturbing fields. Greybody factors serve as a bridge between the near-horizon physics of BHs and asymptotic infinity, quantifying the deviations from perfect blackbody radiation \cite{Visser:1998ke,Okabayashi:2024,Boonserm:2023, Bai:2023,Cox:2023,Oshita:2023}. By analyzing these coefficients, one can gain deeper insights into the wave scattering and the observational imprints of exotic parameters like the symmetry-breaking scale $\eta$ and the Bardeen parameter $b$. This aspect remains underexplored for the Bardeen-like AdS BH with a PGM, particularly when axial and scalar perturbations are considered.

This paper is motivated by the need to fill this gap by extending the study of Bardeen-like AdS BHs to incorporate an in-depth analysis of greybody factors and their implications for astrophysical observations. Our primary aim is to explore how the parameters of the spacetime, including $b$ and $\eta$, influence the geodesic motion of test particles, modify the effective potential governing perturbations, and affect the transmission and reflection probabilities of perturbing waves. By doing so, we aim to shed light on the observational consequences of these modified BHs in astrophysical scenarios and their implications for quantum gravity theories. Specifically, we derive the effective potential for null and time-like geodesics, highlighting the role of PGMs in shaping particle trajectories. We then investigate the RW potential for perturbing fields with different spin states, analyzing the quasinormal spectrum using the sixth-order WKB approximation. Following this, we focus on the computation of greybody factors via the reflection and transmission coefficients for scalar and axial fields, using semi-analytic and numerical methods to derive rigorous bounds. 

The paper is organized as follows: In Sect.~\ref{sec1}, we introduce the metric of the Bardeen-like AdS BH with a PGM, discussing its importance and the key modifications brought by the symmetry-breaking parameter $\eta$ and the Bardeen parameter $b$. Sect.~\ref{sec2} is devoted to the analysis of geodesic motions, where we derive the effective potential for null and time-like particles and explore their trajectories. In Sect.~\ref{sec3}, we investigate the RW potential for different spin fields, analyze the QNM spectrum using the sixth-order WKB approximation, and discuss the stability of the BH under perturbations. Sect.~\ref{sec4} focuses on the computation of greybody factors, analyzing the reflection and transmission probabilities for scalar and axial fields, and their dependence on $\eta$ and $b$. Finally, in Sect.~\ref{sec5}, we present our concluding remarks and highlight potential directions for future research.

\section{\large Bardeen-like AdS BH with phantom global monopoles (PGM)}\label{sec2}

{\color{black} Motivated by the Bardeen regular black hole solution \cite{AA2,CLAR} and the singular black hole with phantom global monopoles \cite{MBAV,SC}, the current study introduces a metric ansatz for a black hole that incorporates phantom global monopoles. This includes a position-dependent mass term analogous to the Bardeen space-time, specifically $\frac{2\,M}{r} \to \frac{2\,M(r)}{r}$, where $M(r)=M\,(1+\mathrm{b}^2/r^2)^{-3/2}$. Thus, the metric ansatz for a static and spherically symmetric Bardeen-like AdS black hole in the presence of phantom global monopoles is described by the following line element in Schwarzschild coordinates $(t, r, \theta, \phi)$ given by}
\begin{eqnarray}
ds^2=-\mathcal{F}(r)\,dt^2+\mathcal{F}(r)^{-1}\,dr^2+r^2\left(d \theta^2+\sin ^2 \theta \,d \phi^2\right),\quad\quad
\mathcal{F}(r)=1-8\,\pi\, \eta^2\, \xi-\frac{2\,M\, r^2}{(r^2+\mathrm{b}^2)^{3/2}}-\frac{\Lambda}{3} r^2,\label{bb1}
\end{eqnarray}
where $M$ is the mass of the BH, $\eta^2$ is the energy scale of the symmetry-breaking, $\Lambda$ is the cosmological constant, $\mathrm{b}$ is the Bardeen-like parameter, and $\xi=1$ (ordinary global monopole) and $\xi=-1$ (phantom global monopole). {\color{black}To derive the metric \eqref{bb1}, we consider the following action:
\begin{equation} \label{izaction}
    S = \int d^4x\, \sqrt{-g}\, \left[ \frac{1}{16\,\pi}\,(R-2\,\Lambda) - \frac{1}{4\,\pi}\, \mathcal{L}(F) - \frac{1}{2}\, \xi\, g^{\mu\nu} (\partial_{\mu} \psi^a)(\partial_{\nu}\psi^a)-\frac{\lambda}{4} (\psi^a\,\psi^a - \eta^2)^2 \right],
\end{equation}
where $R$ is the Ricci scalar; as aforementioned before, $\Lambda$ is the cosmological constant, $\psi^a=\frac{\eta\,f(r)\,x^i}{r}$ is the triplet phantom scalar field with a global symmetry breaking scale $\eta$ and $x^i\,x_i=r^2$ \cite{MBAV,SC}, $\lambda$ is a coupling constant, and $\mathcal{L}(F)$ is the NLED Lagrangian. Further details on the justification and derivation of the field equations are provided in \ref{appendix}.}

In the limit when $\xi =0$, we recover the Bardeen-like AdS BH metric \cite{CLAR}. Moreover, in the limit when $\mathrm{b}=0$, we find a spherically symmetric AdS BH metric with phantom global monopoles \cite{MBAV} which further reduces to the Schwrazschild AdS BH solution for $\xi=0$. We aim to analyze the effect of phantom global monopole on various physics systems.

\begin{figure}[ht!]
    \centering
    \includegraphics[width=0.4\linewidth]{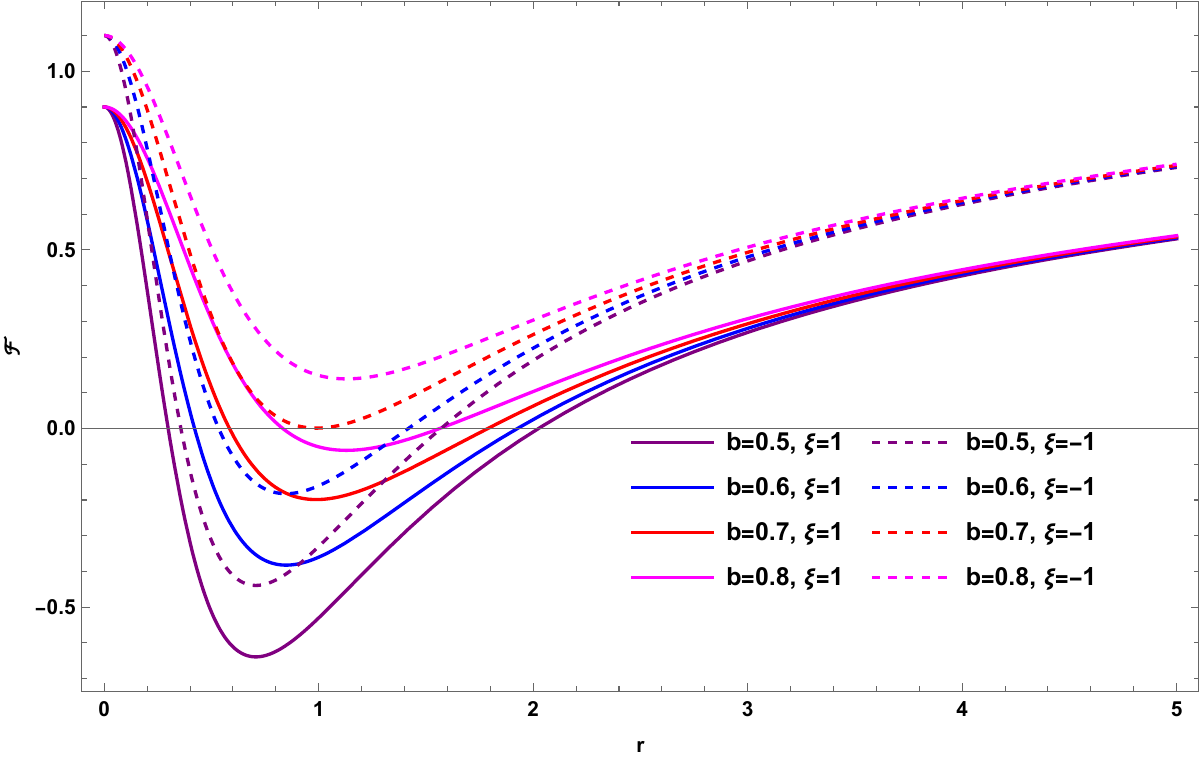}\quad\quad\quad
    \includegraphics[width=0.4\linewidth]{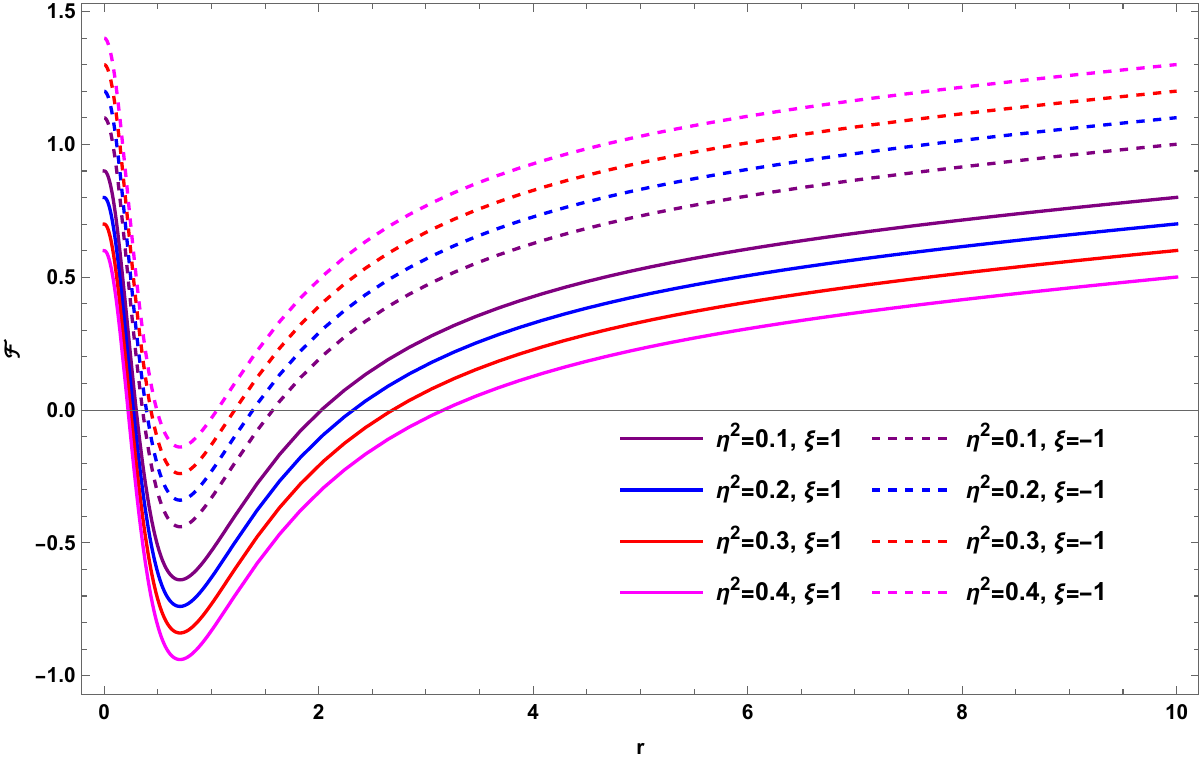}
    \caption{The behavior of the metric function $\mathcal{F}(r)$ for different values of $b$ and $\eta$. The solid lines correspond to GM with $\xi=1$ and dashed lines for phantom GM, $\xi=-1$. Here, we set $M=1$, $\Lambda=-0.003$. In the top panel $\eta^2=0.1$, bottom panel: $b=0.5$.}
    \label{fig:metric-function-1}
\end{figure}

\begin{figure}[ht!]
    \centering
    \includegraphics[width=0.4\linewidth]{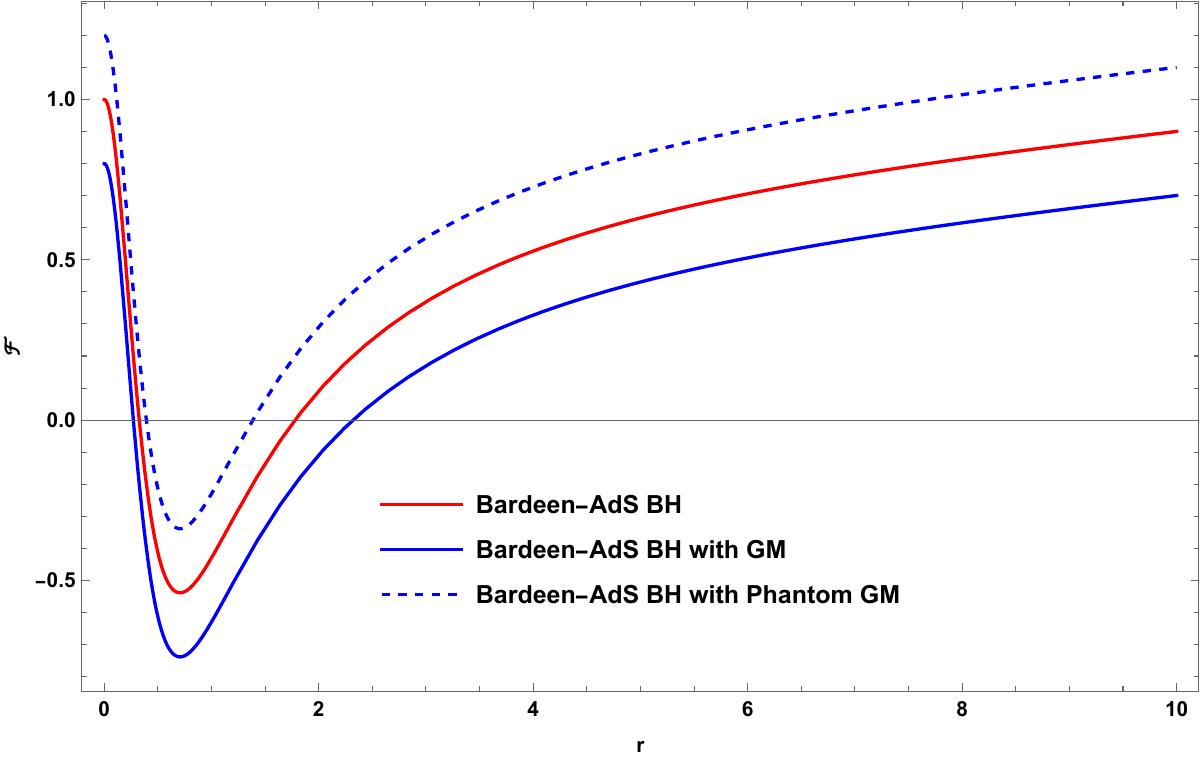}
    \caption{A comparison of the metric function $\mathcal{F}(r)$ for different BHs. Bardeen-AdS BH: $\eta=0$, $\mathrm{b}=0.5$; Bardeen-AdS BH with GM: $\eta^2=0.2$, $\mathrm{b}=0.5$, $\xi=1$, Bardeen-AdS BH with phantom GM: $\eta^2=0.2$, $\mathrm{b}=0.5$, $\xi=-1$.}
    \label{fig:metric-function-2}
\end{figure}

{\color{black}
In Figure \ref{fig:metric-function-1}, we depict the metric function $\mathcal{F}(r)$ as a function of $r$, showing the variation with respect to the energy scale parameter $\eta$ and the Bardeen-like parameter $\mathrm{b}$ for both an ordinary global monopole ($\xi=1$) and a phantom global monopole (($\xi=-1$) in an AdS space, with $\Lambda = -0.003$. The metric function's behavior exhibits noticeable changes as the values of the parameters $\eta$ and $\mathrm{b}$ increase. 

In Figure \ref{fig:metric-function-2}, we compare the metric function $\mathcal{F}(r)$  across different black hole (BH) scenarios. It is evident that, in the presence of an ordinary global monopole, the event horizons is located within the interval $2 < r_{+} < 3$ (in natural units), in contrast to the other scenarios. 

To determine whether the selected black hole (\ref{bb1}) is singular or regular, we calculate the Ricci curvature, $\mathcal{R}=g^{\mu\nu}\,R_{\mu\nu}$, quadratic invariant of Ricci tensor, $\mathcal{R}^2=R_{\mu\nu}\,R^{\mu\nu}$, and  the Kretschmann scalar $\mathcal{K}=R^{\mu\nu\sigma\tau}\,R_{\mu\nu\sigma\tau}$ for the metric (\ref{bb1}). These are given by :
\begin{eqnarray}
    \mathcal{R}&=&\frac{6\,\mathrm{b}^2\, M\,(4\,\mathrm{b}^2-r^2)}{(\mathrm{b}^2 + r^2)^{7/2}}+\frac{2\,\xi\,\eta^2}{r^2}+4\,\Lambda\quad (\Lambda \neq 0),\label{ricci1}\\
    &=&\frac{6\,\mathrm{b}^2\, M\,(4\,\mathrm{b}^2-r^2)}{(\mathrm{b}^2 + r^2)^{7/2}}+\frac{2\,\xi\,\eta^2}{r^2}\quad (\Lambda = 0).\label{ricci2}
\end{eqnarray}
\begin{eqnarray}
    \mathcal{R}^2&=&\frac{450\, \mathrm{b}^8\, M^2}{(\mathrm{b}^2 + r^2)^7}-\frac{540\, \mathrm{b}^6\, M^2}{(\mathrm{b}^2 + r^2)^6}+4\,\Lambda^2+6\,M\,\mathrm{b}^4\,\left(\frac{39\, M}{(\mathrm{b}^2 + r^2)^5}+\frac{10\,\Lambda}{(\mathrm{b}^2 + r^2)^{7/2}}\right)+\frac{4\,\xi\,\eta^2\,\Lambda}{r^2}+\frac{2\,\eta^4\,\xi^2}{r^4}\nonumber\\
    &-&\frac{12\, \mathrm{b}^2\, M\,(r^2\,\Lambda-2\,\eta^2\,\xi)}{r^2\,(\mathrm{b}^2 + r^2)^{5/2}}\quad (\Lambda \neq 0),\label{qua1}\\
    &=&\frac{9\,M^2\,(2\, \mathrm{b}^4 - 3\, \mathrm{b}^2\, r^2)^2}{(\mathrm{b}^2 + r^2)^7}+\frac{9\,(2\, \mathrm{b}^4\, M - 3\, \mathrm{b}^2\, M\, r^2)^2}{(\mathrm{b}^2 + r^2)^7}+2\,\left(\frac{6\, \mathrm{b}^2\, M}{(\mathrm{b}^2 + r^2)^{5/2}}+ \frac{\eta^2\,\xi}{r^2}\right)^2\quad (\Lambda = 0).\label{qua2}
\end{eqnarray}
\begin{eqnarray}
    \mathcal{K}&=&\frac{4}{3}\,\Bigg[\frac{6\,M}{r^2\, (\mathrm{b}^2 + r^2)^{3/2}}+\frac{9\, M^2\, (8\, \mathrm{b}^8 - 4\, \mathrm{b}^6\, r^2 + 47\, \mathrm{b}^4\, r^4 - 12\, \mathrm{b}^2\, r^6 + 4\, r^8)}{(\mathrm{b}^2 + r^2)^7}+2\,\Lambda^2+\frac{2\,\Lambda\,\eta^2\,\xi}{r^2}+\frac{3\,\eta^4\,\xi^2}{r^4}\nonumber\\
    &&+\frac{6\, M\,\left\{-\mathrm{b}^2\, r^2\, (2 + r^2\, \Lambda - 4\, \eta^2\, \xi)+r^4\,(-1 + 2\, \eta^2\, \xi)+\mathrm{b}^4\, (-1 + 4\, r^2\, \Lambda+2\,\eta^2\,\xi)\right\}}{r^2\, (\mathrm{b}^2 + r^2)^{7/2}} \Bigg]\quad (\Lambda \neq 0),\label{krets1}\\
    &=&4\,\Bigg[\frac{M^2\,\left\{4\,(-2\,\mathrm{b}^2+r^2)\,(\mathrm{b}^2+r^2)^2+(2\, \mathrm{b}^4 -11\, \mathrm{b}^2\, r^2 + 2\, r^4)^2\right\}}{(\mathrm{b}^2 + r^2)^7}+\left(\frac{2\,M}{(\mathrm{b}^2 + r^2)^{3/2}}+\frac{\eta^2\,\xi}{r^2}\right)^2\Bigg]\quad (\Lambda = 0).\label{krets2}
\end{eqnarray}
One can easily check that the Ricci scalar (\ref{ricci1}), the Ricci quadratic invariant (\ref{qua1}) and the Kretschmann scalar (\ref{krets1}) diverge as $r \to 0$, even in the case of a zero cosmological constant. This indicates the presence of a central singularity at the origin. However, as $r \to \infty$, these scalar approaches $4\,\Lambda$, $4\,\Lambda^2$ and $\frac{8}{3}\,\Lambda^2$, respectively, consistent with an asymptotically AdS space. Therefore, we can conclude that the chosen space-time (\ref{bb1}) represents an example of a singular black hole with phantom global monopoles.
}

\subsection{\large \bf Thermodynamic Properties of BH}

{\color{black} In this section, we examine the thermal properties associated with the selected black hole solution (\ref{bb1}), where the cosmological constant $\Lambda$ is treated as a dynamical variable pressure, denoted by $P$. This interpretation can be justified by the Smarr relation and the first law of black hole thermodynamics in asymptotically AdS space-times. The thermodynamic pressure $P$ is related to the cosmological constant $\Lambda$ through the equation: $P=-\frac{\Lambda}{8\,\pi}$. To determine the black hole mass $M$, we apply the condition $\mathcal{F}(r_{+}) = 0$ at the event horizon. This mass is given by
\begin{equation}
    M=\frac{(r^2_{+}+\mathrm{b}^2)^{3/2}\,\left(1-8\,\pi\, \eta^2\, \xi-\frac{\Lambda}{3} r^2_{+}\right)}{2\,r^2_{+}}.\label{mass}
\end{equation}
}

The Hawking temperature \cite{Hawking:1975, Bekenstein:1973, Sakalli:2012, Page:2005,NPB} of the selected black hole is obtained as,
\begin{equation}
T_H = \frac{1}{4\pi} \left. \frac{d\mathcal{F}(r)}{dr} \right|_{r=r_+}.
\end{equation}
Substituting Eq. (\ref{bb1}), we obtain:
\begin{equation}
T_H = \frac{1}{4\,\pi} \left[ -\frac{6\, M\, r_+}{(r_+^2 + \mathrm{b}^2)^{3/2}} + \frac{6\, M\, r_+^3}{(r_+^2 + \mathrm{b}^2)^{5/2}} - \frac{2\,\Lambda\, r_+}{3} \right].
\end{equation}

The Bekenstein-Hawking entropy \cite{Bekenstein:1973, Hawking:1975,Strominger:1996} is given by:
\begin{equation}
S = \frac{A}{4} = \pi r_+^2.
\end{equation}

The heat capacity at constant pressure \cite{Johnson:2024} is computed as:
\begin{equation}
C_P = \left( \frac{\partial M}{\partial T_H} \right)_{P}.
\end{equation}
After computation, the explicit form of $C_P$ is found to be:
\begin{equation}
C_P = 2\,\pi\, r_+ \left[ \frac{6\, M\, r_+ \mathrm{b}^2 (r_+^2 + \mathrm{b}^2)^{-5/2} + \frac{2\,\Lambda\, r_+}{3}}{6\, M\, (r_+^2 - 2\,\mathrm{b}^2) (r_+^2 + \mathrm{b}^2)^{-5/2} - \frac{2\,\Lambda}{3}} \right].
\end{equation}
The sign of $C_P$ determines the thermodynamic stability of the BH, with $C_P > 0$ indicating local stability and $C_P < 0$ suggesting instability. {\color{black}To analyze whether $C_P > 0$, we examine the numerator and denominator separately. The numerator remains positive for physical values of $M, r_+$, and $b$, whereas the denominator varies depending on $r_+$. For small $r_+$, the dominant term in the denominator is $6\, M\, (r_+^2 - 2\,\mathrm{b}^2) (r_+^2 + \mathrm{b}^2)^{-5/2}$. Since $r_+^2 - 2\mathrm{b}^2$ can be negative for sufficiently small $r_+$, the denominator also becomes negative, leading to $C_P < 0$ and indicating thermodynamic instability. For large $r_+$, the term $\frac{2\Lambda}{3}$ in the denominator, which is negative due to $\Lambda < 0$ in AdS space, further decreases the denominator, maintaining the negativity of $C_P$ and thus suggesting persistent instability. However, an intermediate range of $r_+$ may exist where the competing effects of the $6M$ term and the $\frac{2\Lambda}{3}$ term in the denominator result in a positive $C_P$, allowing for a possible stable phase (see Figure \ref{izzet}).

Thus, unlike asymptotically flat cases where stability emerges at large $r_+$, the negative cosmological constant in AdS space complicates the stability conditions. The system may exhibit a phase transition at a critical $r_+$ where $C_P$ changes sign, beyond which the thermodynamic behavior is dictated by the dominance of the cosmological term. 

\begin{figure}[ht!]
   \centering 
   \includegraphics[width=.8\textwidth]{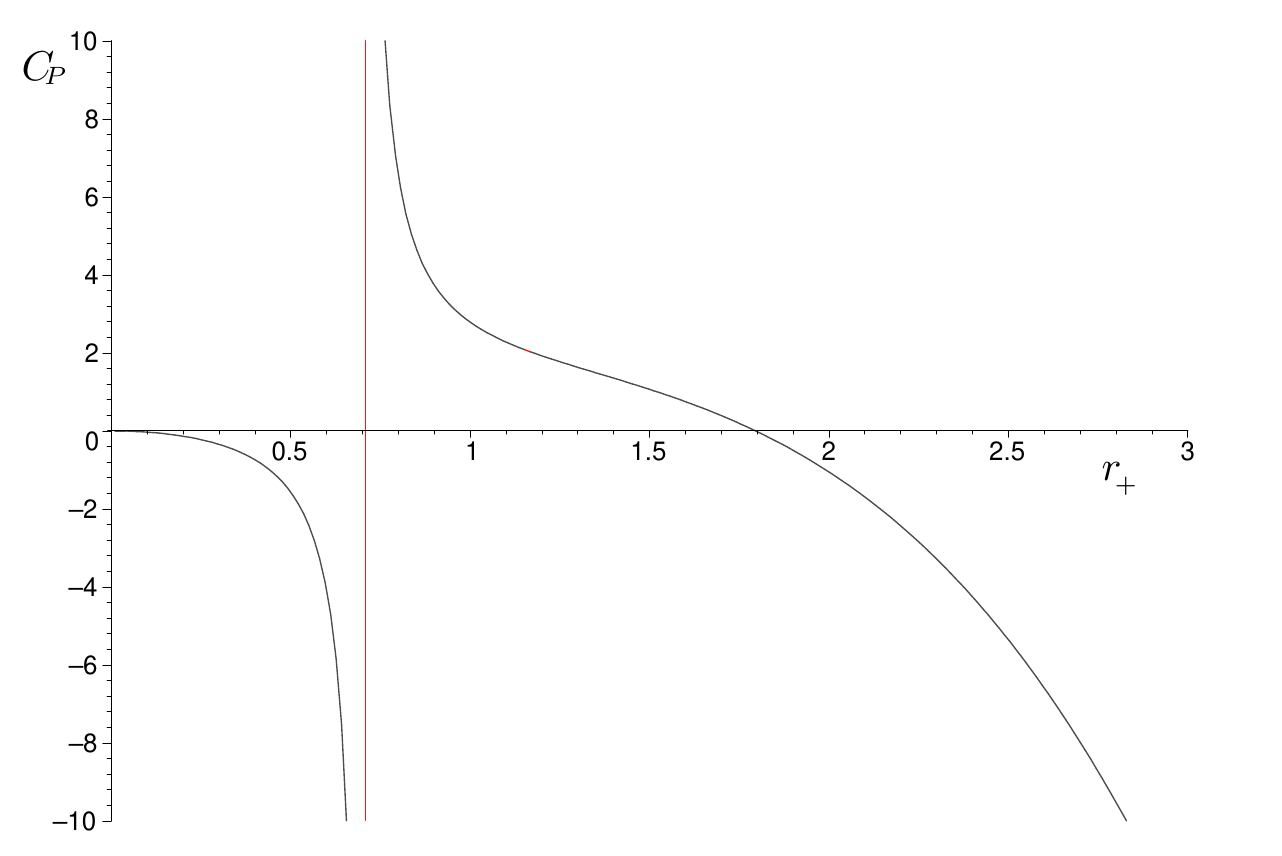}\quad\quad
  \caption{{\color{black} Plot of the heat capacity $C_P$ as a function of the black hole horizon radius $r_+$. The vertical red line represents a divergence point, indicating a phase transition. The plot shows regions where $C_P > 0$ (thermodynamically stable) and $C_P < 0$ (thermodynamically unstable), suggesting a transition between stable and unstable black hole phases. The physical parameters are chosen as $M=1$, $b=0.5$, and $\Lambda=-0.1$.}} \label{izzet}
\end{figure}
}
\subsection{\bf Geodesic Motions: Null and Time-like particles}

{\color{black} In this section, we investigate the geodesic motion of both massive and massless particles in the gravitational field produced by the black hole (\ref{bb1}), examining how the presence of phantom global monopoles and other parameters influences the dynamics of the particles. We adopt a general formalism known as the Lagrangian approach to model and analyze the motion of test particles in the immediate vicinity of the selected black hole space-time, as described by the line element in Eq. (\ref{bb1}). This approach has been widely employed by numerous authors to study the motion of massless and massive test particles in curved space-times (see, Refs. \cite{NPB,CJPHY,EPJC,Ahmed:2024mnz,AHEP1,AHEP2,AHEP3,AHEP4}).} 

The Lagrangian density function of a system is defined by
\begin{equation}
    \mathcal{L}=\frac{1}{2}\,g_{\mu\nu}\,\dot{x}^{\mu}\,\dot{x}^{\nu},\label{bb2}
\end{equation}
where dot represent ordinary derivative w. r. t. an affine parameter $\tau$ and $g_{\mu\nu}$ is the metric tensor.

{\color{black} Given the spherically symmetric nature of the black hole spacetime (\ref{bb1}), we further consider the geodesic motion of particles restricted to the equatorial plane, defined by $\theta = \pi/2$. It is well known that the Lagrangian density function does not explicitly depend on the coordinates $t$ and $\phi$, which leads to the definition of constants of motion, namely $p_t=-\mathrm{E}$ and $p_{\phi}= \mathrm{L}$. Here, $\mathrm{E}$ and $\mathrm{L}$ correspond to the energy and angular momentum of the particles, respectively. The equations of motion can be derived as follows:}
\begin{eqnarray}
    \dot{t}=\frac{\mathrm{E}}{\mathcal{F}(r)},\quad\quad\quad
    \dot{\phi}=\frac{\mathrm{L}}{r^2},\quad\quad\quad
    \dot{r}^2=\mathrm{E}^2-V_\text{eff},\label{bb3}
\end{eqnarray}
where $V_\text{eff}$ is the effective potential of the system. 

\begin{figure*}[ht!]
  \includegraphics[width=.45\textwidth]{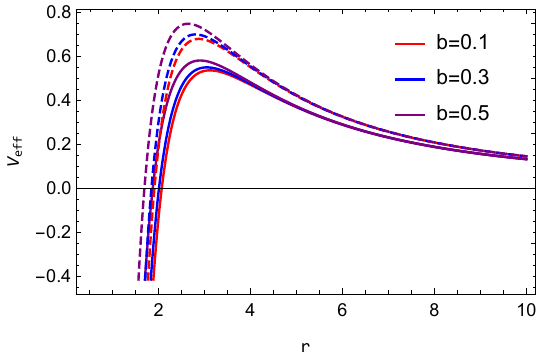}\quad\quad
  \includegraphics[width=.45\textwidth]{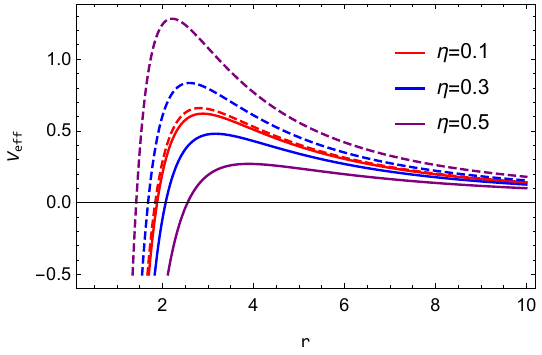}
\caption{Null geodesic effective potential affected by Bardeen parameter $b$ (left) and the energy scale of the symmetry breaking $\eta$ (right). Here $\xi = 1$ solid, $\xi=-1$ dashed.} \label{FS2}
\end{figure*}

For null or time-like geodesics, this effective potential is given by
\begin{equation}
    V_\text{eff}=\left(-\epsilon+\frac{\mathrm{L}^2}{r^2}\right)\,\left[1-8\,\pi\, \eta^2\, \xi-\frac{2\,M\, r^2}{(r^2+\mathrm{b}^2)^{3/2}}-\frac{\Lambda}{3} r^2\right].\label{bb4}
\end{equation}
{\color{black} Here $\epsilon=0$ for null geodesics and $-1$ for time-like.

From the expression in Eq. (\ref{bb4}), it is evident that the effective potential for null or time-like geodesics is influenced by several factors (see Fig. \ref{FS2}). These include the Bardeen-like parameter $\mathrm{b}$, the energy scale parameter $\eta$, and the cosmological constant $\Lambda$ for a specific angular momentum $\mathrm{L}$. In comparison to the case of an ordinary global monopole ($\xi=1$), the influence of a phantom global monopole ($\xi=-1$) results in a more pronounced modification of the effective potential. This indicates that the nature of the global monopole significantly alters the dynamics of test particles, as well as their motion in the gravitational field produced by the black hole. The inclusion of these parameters not only changes the shape of the effective potential but also influences the behavior of both massive and mass-less particles in the vicinity of the black hole.} 

The behaviors of the effective potential due to the geodesics equation, let's call it geodesics effective potential $V_\text{eff}$, are drawn in Fig. \eqref{FS2} for Bardeen parameter $b$ (left) and energy scale $\eta$ (right) parameters. 

\begin{center}
    \large{\bf I.\,\,Null Geodesics: Motions of Photon Light}
\end{center}

To study the behaviour photon light under the influence of the gravitational field, the effective potential for null geodesics from Eq. (\ref{bb4}) becomes
\begin{eqnarray}
    V_\text{eff}=\frac{\mathrm{L}^2}{r^2}\,\left[1-8\,\pi\, \eta^2\, \xi-\frac{2\,M\, r^2}{(r^2+\mathrm{b}^2)^{3/2}}-\frac{\Lambda}{3} r^2\right].\label{bb4a}
\end{eqnarray}

For circular null geodesics, we have the conditions $\dot{r}=0$ and $\ddot{r}=0$ at $r=r_c$. These implies from Eq. (\ref{bb3}) that
\begin{equation}
    V_\text{eff}(r_c)=\mathrm{E}^2\Rightarrow \frac{1}{\beta^2_c}=\frac{1-8\,\pi\, \eta^2\, \xi}{r^2_{c}}-\frac{2\,M}{(r^2_{c}+\mathrm{b}^2)^{3/2}}-\frac{\Lambda}{3},\quad\quad \beta_c=\mathrm{L}/\mathrm{E}.\label{bb4b}
\end{equation}
And
\begin{equation}
    V'_\text{eff}(r=r_c)=0\Rightarrow 2\,\mathcal{F}(r_c)=r_c\,\mathcal{F}'(r_c).\label{bb4c}
\end{equation}

{\color{black} The principal Lyapunov exponents null geodesics can be expressed as \cite{VC} ( (note that
in Ref. \cite{VC}, the authors used metric signature $-2$ while in our case it will be $+2$)
\begin{equation}
    \lambda^\text{null}_L=\sqrt{-\frac{V''_\text{eff}(r_c)}{2\,\dot{t}^2}}.\label{lp}
\end{equation}
Thereby, substituting Eqs. (\ref{bb3}), (\ref{bb4b}) and (\ref{bb4c}) into the Eq. (\ref{lp}), for circular null geodesics we find
\begin{equation}
    \lambda^\text{null}_L=\sqrt{\frac{\mathcal{F}(r_c)}{r^2_{c}}\,\left(\frac{\mathcal{F}(r_c)}{r_c}-\frac{\mathcal{F}''(r_c)}{2}\right)},\quad\quad V''_\text{eff}(r_c)=\frac{\mathrm{L}^2}{r^4_{c}}\,\left(r_c\,\mathcal{F}''(r_c)-2\,\mathcal{F}(r_c)\right).\label{lp2}
\end{equation}

Substituting the metric function $\mathcal{F}$ given in Eq. (\ref{bb1}), we find the following expression for the Lyapunov exponent as, {\small
\begin{eqnarray}
    \lambda^\text{null}_L=\frac{1}{r_c}\sqrt{\frac{1}{3}\left(1-8\,\pi\, \eta^2\, \xi-\frac{2\,M\, r^2_c}{(r^2_c+\mathrm{b}^2)^{3/2}}-\frac{\Lambda}{3}\,r^2_c\right)\left(\frac{45\,M\,r^4_c}{(r^2_c+\mathrm{b}^2)^{7/2}}-\frac{45\,M\,r^2_c}{(r^2_c+\mathrm{b}^2)^{5/2}}-\frac{6\,M\,r_c}{(r^2_c+\mathrm{b}^2)^{3/2}}+(1-r_c)\,\Lambda+\frac{3\,(1-8\,\pi\,\eta^2\,\xi)}{r_c}\right)}.\label{lp3}
\end{eqnarray}
}

The expression for the Lyapunov exponent shows that it is influenced by several factors, including the energy scale $\eta$, the Bardeen-like parameter $\mathrm{b}$, the cosmological constant $\Lambda$. Additionally, it is observed that the Lyapunov exponent is higher in the presence of a phantom global monopole ($\xi=-1$) compared to an ordinary global monopole ($\xi=1$).

An important quantity for the analysis of the null geodesics is the angular frequency ($\Omega_c$) or the coordinate angular velocity at the null geodesic given by \cite{VC}
\begin{equation}
    \Omega_c=\frac{\dot{\phi}}{\dot{t}}=\sqrt{\frac{\mathcal{F}'(r_c)}{2\,r_c}}.\label{lp4}
\end{equation}
In our study, using (\ref{bb4c}) we find the following expression
\begin{equation}
    \Omega_c=\frac{\sqrt{\mathcal{F}(r_c)}}{r_c}=\frac{1}{r_c}\,\sqrt{1-8\,\pi\, \eta^2\, \xi-\frac{2\,M\, r^2_c}{(r^2_c+\mathrm{b}^2)^{3/2}}-\frac{\Lambda}{3}\,r^2_c}.\label{lp5}
\end{equation}
The expression for the angular frequency shows that it is influenced by several factors, including the energy scale $\eta$, the Bardeen-like parameter $\mathrm{b}$, the cosmological constant $\Lambda$. Additionally, it is observed that the angular frequency is higher in the presence of a phantom global monopole ($\xi=-1$) compared to an ordinary global monopole ($\xi=1$).
}

Now, we find photon trajectory equation in the gravitational field. From Eq. (\ref{bb3}), we define the following quantity
\begin{eqnarray}
    \frac{\dot{r}^2}{\dot{\phi}^2}=\left(\frac{dr}{d\phi}\right)^2&=&r^4\,\Bigg[\frac{\mathrm{E}^2}{\mathrm{L}^2}+\frac{\Lambda}{3}-\frac{1-8\,\pi\, \eta^2\, \xi}{r^2}+\frac{2\,M}{(r^2_{c}+\mathrm{b}^2)^{3/2}}\Bigg].\quad\label{bb4d}
\end{eqnarray}
Transforming to a new variable $u=1/r$ into the above equation yields
\begin{eqnarray}
    \left(\frac{du}{d\phi}\right)^2+(1-8\,\pi\, \eta^2\, \xi)\,u^2=\frac{1}{\beta^2}+\frac{\Lambda}{3}+\frac{2\,M\,u^3}{(1+u^2\,\mathrm{b}^2)^{3/2}},\label{bb4e}
\end{eqnarray}
where $\beta=\mathrm{L}/\mathrm{E}$ is the impact parameter for photon light. 

Equation (\ref{bb4e}) is the photon light trajectory under the influence of the gravitational field produced by the selected BH solution. From the above photon trajectory equation, it is evident that photon light paths is influenced by a phantom ($\xi=-1$) and an ordinary ($\xi=1$) global monopole, and thus, deviates from the original path in the case of standard BH.  

Finally, we focus on photon sphere and determine its size. The photon sphere radius can be calculated using the definition $V'_\text{eff}=0$ at $r=r_\text{ph}$ which implies the following relation
\begin{eqnarray}
    x^{3}+x^{8}\,\mathrm{b}^2=\mathcal{M}^2,\quad\quad x=r^{1/5}_\text{ph},\quad\quad \mathcal{M}=\left(\frac{3\,M}{1-8\,\pi\,\eta^2\,\xi}\right)^{1/5}.\label{bb7}
\end{eqnarray}

Finally, we determine force on the massless photon particles under the influence of the gravitational field produced by the selected AdS background BH and analyze the outcomes. This force can be defined in terms of the effective potential for null geodesics as $\mathrm{F}_\text{ph}=-V'_\text{eff}/2$. Using the effective potential expression Eq. (\ref{bb4a}), we find
\begin{equation}
    \mathrm{F}_\text{ph}=\frac{\mathrm{L}^2}{r^3}\,\left[1-8\,\pi\,\eta^2\,\xi-\frac{3\,M\,r^3}{(r^2+\mathrm{b}^2)^{5/2}}\right].\label{force}
\end{equation}

From the expression for the force in Eq. (\ref{force}) acting on photon light, it is evident that the presence of phantom global monopoles modifies this force, leading to deviations from the results typically observed in a standard black hole. Additionally, we find that the force on photon light is stronger in the presence of a phantom global monopole ($\xi=-1$) compared to an ordinary global monopole ($\xi=1$) within the specific black hole metric.

To facility the photon sphere, we will use the following tables. 
\begin{table}[ht!]
\centering
\caption{Ordinary global monopole, {\color{black} $\xi=1$} (setting $8\,\pi=1$)}\label{tab:1} 
\begin{tabular}{|l|l|l|l|l|l|}
   \hline
    $\eta^2$ & $0$ & $0.1$ & $0.2$ & $0.3$ & $0.4$ \\
    \hline
     ${\color{black}\mathrm{a}=1-\eta^2}$ &  $1$ & $0.9$ & $0.8$ & $0.7$ & $0.6$ \\ [0.5ex] 
     \hline
\end{tabular}
\hfill\\
\centering
\caption{Phantom global monopole, {\color{black}$\xi=-1$} (setting $8\,\pi=1$)}\label{tab:2}
\begin{tabular}{|l|l|l|l|l|l|l|}
    \hline
    $\eta^2$ & $0$ & $0.10$ & $0.12$ & $0.14$ & $0.16$ & $0.18$ \\ 
    \hline
     ${\color{black}\mathrm{a'}=1+\eta^2}$ &  $1$ & $1.1$ & $1.12$ & $1.14$ & $1.16$ & $1.18$ \\ 
     \hline
\end{tabular}
\end{table} 

\begin{center}
    \large{\bf II.\,\, Time-like Geodesics: Motions of Massive Particles}
\end{center}

In this part, we study dynamics of time-like particles around the selected BH and analyze the outcomes. For time-like geodesics, the effective potential from Eq. (\ref{bb4}) becomes
\begin{eqnarray}
    V_\text{eff}=\left(1+\frac{\mathrm{L}^2}{r^2}\right)\,\left[1-8\,\pi\, \eta^2\, \xi-\frac{2\,M\, r^2}{(r^2+\mathrm{b}^2)^{3/2}}-\frac{\Lambda}{3} r^2\right].\label{fff1}
\end{eqnarray}

In the equatorial plane, for circular orbits of time-like particles, the conditions $\dot{r}=0$ and $\ddot{r}=0$ hold true. Consequently, the energy of the time-like particles is expressed as
\begin{eqnarray}
    \mathrm{E}_{\pm}=\pm\,\frac{\left(1-8\,\pi\, \eta^2\, \xi-\frac{2\,M\, r^2}{(r^2+\mathrm{b}^2)^{3/2}}-\frac{\Lambda}{3} r^2\right)}{\sqrt{1-8\,\pi\,\eta^2\,\xi-\frac{3\,M\,r^4}{(r^2+\mathrm{b}^2)^{5/2}}}},\label{fff2}
\end{eqnarray}
The angular momentum for circular orbits is described by
\begin{equation}
    \mathrm{L}=\sqrt{\frac{r^2\,\left(\frac{M\, r^2\,(r^2-2\,\mathrm{b}^2)}{(r^2+\mathrm{b}^2)^{3/2}}-\frac{\Lambda}{3}\,r^2\right)}{1-8\,\pi\,\eta^2\,\xi-\frac{3\,M\,r^4}{(r^2+\mathrm{b}^2)^{5/2}}}}.\label{fff3}
\end{equation}

It is evident from Eqs. (\ref{fff2}) and (\ref{fff3}) that the energy and angular momentum of time-like particles on equatorial circular orbits are affected by both phantom and ordinary global monopoles, leading to deviations from the standard BH scenario.

The orbital angular velocity of time-like particles on the circular orbits in the equatorial plane is given by
\begin{equation}
    \Omega=\sqrt{\frac{\mathcal{F}'(r)}{2\,r}}=\sqrt{\frac{M\,(r^2-2\,\mathrm{b}^2)}{(r^2+\mathrm{b}^2)^{5/2}}-\frac{\Lambda}{3}}.\label{fff4}
\end{equation}

Subsequently, we compute the Lyapunov exponent to assess the stability or instability of equatorial time-like orbits around the BH, expressed via the effective potential (\ref{fff1}) as \cite{VC}
\begin{equation}
    \lambda^\text{timelike}_{\mathrm{L}}=\sqrt{\frac{-V''_\text{eff}}{2\,\dot{t}^2}}=\sqrt{\frac{2\,(\mathcal{F}'(r))^2-\mathcal{F}(r)\,\mathcal{F}''(r)-\frac{3\,\mathcal{F}(r)\,\mathcal{F}'(r)}{r}}{2}}.\label{fff5}
\end{equation}
Using the metric function (\ref{bb1}), we find the following expression of the Lyapunov exponent as, {\small
\begin{eqnarray}
    \lambda^\text{timelike}_{\mathrm{L}}=\left[\frac{6\,M^2\,r^6}{(r^2 +\mathrm{b}^2)^5}- 
 \frac{4\,\Lambda}{3}\,(-1 + \eta^2\,\xi)-\frac{M\,\Big\{(1-\eta^2\,\xi)\,\left(-8\,\mathrm{b}^4 + 8\,\mathrm{b}^2\, r^2 +r^4\right)+5\,r^6\, \Lambda\Big\}}{(r^2 +\mathrm{b}^2)^{7/2}}\right]^{1/2}.\label{fff6}
\end{eqnarray}
}
{\color{black} Considering that $(2\,\mathcal{F}-r\,\mathcal{F}')>0$ and that unstable orbits are characterized by $V''_\text{eff}<0$, we conclude that for unstable circular orbits, the Lyapunov exponent is real, with $\lambda_{\mathrm{L}}>0$. In contrast, for stable orbits, the Lyapunov exponent takes on a complex value \cite{VC}.}  

Finally, we determine the time-like particle trajectory under the influence of the gravitational field and show how various parameters influence the time-like particle path. We defined the following quantity for time-like geodesics using Eq. (\ref{bb3}) as,
\begin{eqnarray}
    \frac{\dot{r}^2}{\dot{\phi}^2}=r^4\,\Bigg[\frac{\mathrm{E}^2}{\mathrm{L}^2}-\left(\frac{1}{\mathrm{L}^2}+\frac{1}{r^2}\right)\left(1-8\,\pi\, \eta^2\, \xi-\frac{2\,M\, r^2}{(r^2+\mathrm{b}^2)^{3/2}}-\frac{\Lambda}{3} r^2\right)\Bigg].\label{fff7}
\end{eqnarray}
Transforming to a new variable via $u(\phi)=1/r(\phi)$ into the above equation results
\begin{eqnarray}
    \left(\frac{du}{d\phi}\right)^2&=&\frac{\mathrm{E}^2}{\mathrm{L}^2}-\left(\frac{1}{\mathrm{L}^2}+u^2\right)
    \left(1-8\,\pi\, \eta^2\, \xi-\frac{2\,M\,u}{(1+\mathrm{b}^2\,u^2)^{3/2}}-\frac{\Lambda}{3\,u^2}\right)\label{fff8}
\end{eqnarray}
where $\mathrm{E}$ and $\mathrm{L}$ are given in Eqs. (\ref{fff2}) and (\ref{fff3}), respectively. 

Equation (\ref{fff7}) is the time-like particles trajectory under the influence of the gravitational field produced by the selected BH. From the above trajectory equation, it is evident that the time-like particle path is influenced by a phantom and an ordinary global monopole, and thus deviates from the original path in the case of standard BH.

\section{\large Perturbations of Bardeen-like AdS BH with PGM: The Regge-Wheeler potential} \label{sec3}

In this section, we investigate the spin-dependent Regge-Wheeler (RW) potentials for the black hole (BH) solution (\ref{bb1}). The RW-Zerilli equations describe the gravitational perturbations of a Schwarzschild-like BH in General Relativity (GR). These perturbations are classified into two categories: axial and polar. The equation governing axial perturbations is the RW equation \cite{Fiziev:2007es}, while the Zerilli equation \cite{Chandrasekhar:1975} governs the polar perturbations. Axial perturbations tend to be more intricate and resist simple treatment through the WKB approximation, making the computation of QNMs challenging without numerical methods. As a result, we focus on examining the RW potential for fields with different spin values, such as spin-zero, spin-one, and spin-two, and analyze the corresponding outcomes. The Regge-Wheeler potential has been extensively studied in the context of various BH spacetimes (see, for example, Refs. \cite{NPB,CJPHY,EPJC} and the references therein). 

To proceed for the RW-potential, we perform the following coordinate change (called tortoise coordinate) 
\begin{eqnarray}
    &&dr_*=\frac{dr}{\mathcal{F}(r)}\label{cc1}
\end{eqnarray}
into the line-element Eq. (\ref{bb1}) results
\begin{equation}
    ds^2=\mathcal{F}(r_*)\,\{-dt^2+dr^2_{*}\}+\mathcal{H}^2(r_*)\,(d\theta^2+\sin^2 \theta\,d\phi^2),\label{cc2}
\end{equation}

In Regge and Wheeler’s original work \cite{Regge:1957}, they show that for perturbations in a BH spacetime, assuming a separable wave form of the type
\begin{equation}
    \Phi(t, r_{*},\theta, \phi)=\exp(i\,\omega\,t)\,Y^{m}_{\ell} (\theta,\phi)\,\psi(r_*)/r_{*},\label{cc3}
\end{equation}
where $Y^{m}_{\ell} (\theta,\phi)$ are the spherical harmonics, $\omega$ is (possibly complex) temporal frequency in the Fourier domain \cite{Poisson:2004}, and $\psi (r)$ is a propagating scalar, vector, or spin two axial bi-vector field in the candidate space-time. RW-equation is given by
\begin{equation}
    \frac{\partial^2 \psi(r_*)}{\partial r^2_{*}}+\left\{\omega^2-\mathcal{V}_S\right\}\,\psi(r_*)=0.\label{cc4}
\end{equation}
The method for solving Equation (\ref{cc4}) is dependent on the spin of the perturbations and on the background space-time.

The spin-dependent RW-potential is given by the following expression
\begin{equation}
    \mathcal{V}_S=\frac{\mathcal{F}}{\mathcal{H}^2}\,\Big\{\ell\,(\ell+1)+S\,(S-1)\,(g^{rr}-1)\Big\}+(1-S)\,\frac{\partial^2_{r_{*}}\,\mathcal{H}}{\mathcal{H}},\label{cc5}
\end{equation}
where $\mathcal{F}$ and $\mathcal{H}$ are the relevant functions as specified by Equation (\ref{bb1}), $\ell_0$ is the multipole number $\ell_0 \geq S$, and $g^{rr}$ is the relevant contrametric component.

For the BH under considerations, we have
\begin{equation}
    \mathcal{H}=r,\quad \partial_{r_*}=\mathcal{F}(r)\,\partial_r.\label{cc6}
\end{equation}

Thus, the spin-dependent RW-potential for the selected BH becomes {\small
\begin{eqnarray}
    \mathcal{V}_S=\left(1-8\pi\eta^2\xi-\frac{2Mr^2}{(r^2+\mathrm{b}^2)^{3/2}}-\frac{\Lambda}{3} r^2\right)\Bigg[\frac{\ell(\ell+1)}{r^2}+\frac{S(S-1)}{r^2}\left(-8\pi\eta^2\xi-\frac{2Mr^2}{(r^2+\mathrm{b}^2)^{3/2}}-\frac{\Lambda}{3} r^2\right)
    +\frac{(1-S)}{r}\left(\frac{2M(r^2-2\mathrm{b}^2)}{(r^2+\mathrm{b}^2)^{5/2}}-\frac{2\Lambda}{3}\right) \Bigg].\label{cc7}
\end{eqnarray}
}

{\color{black}
The expression for the spin-dependent RW-potential (\ref{cc7}) shows that it is influenced by several factors, including the energy scale $\eta$, the Bardeen-like parameter $\mathrm{b}$, the cosmological constant $\Lambda$. Moreover, the orbital quantum number $\ell$ also alters this RW-potential.   Additionally, it is observed that the RW-potential is higher in the presence of a phantom global monopole ($\xi=-1$) compared to an ordinary global monopole ($\xi=1$).
}

\begin{figure*}[ht!]
    \centering
  \includegraphics[width=.45\textwidth]{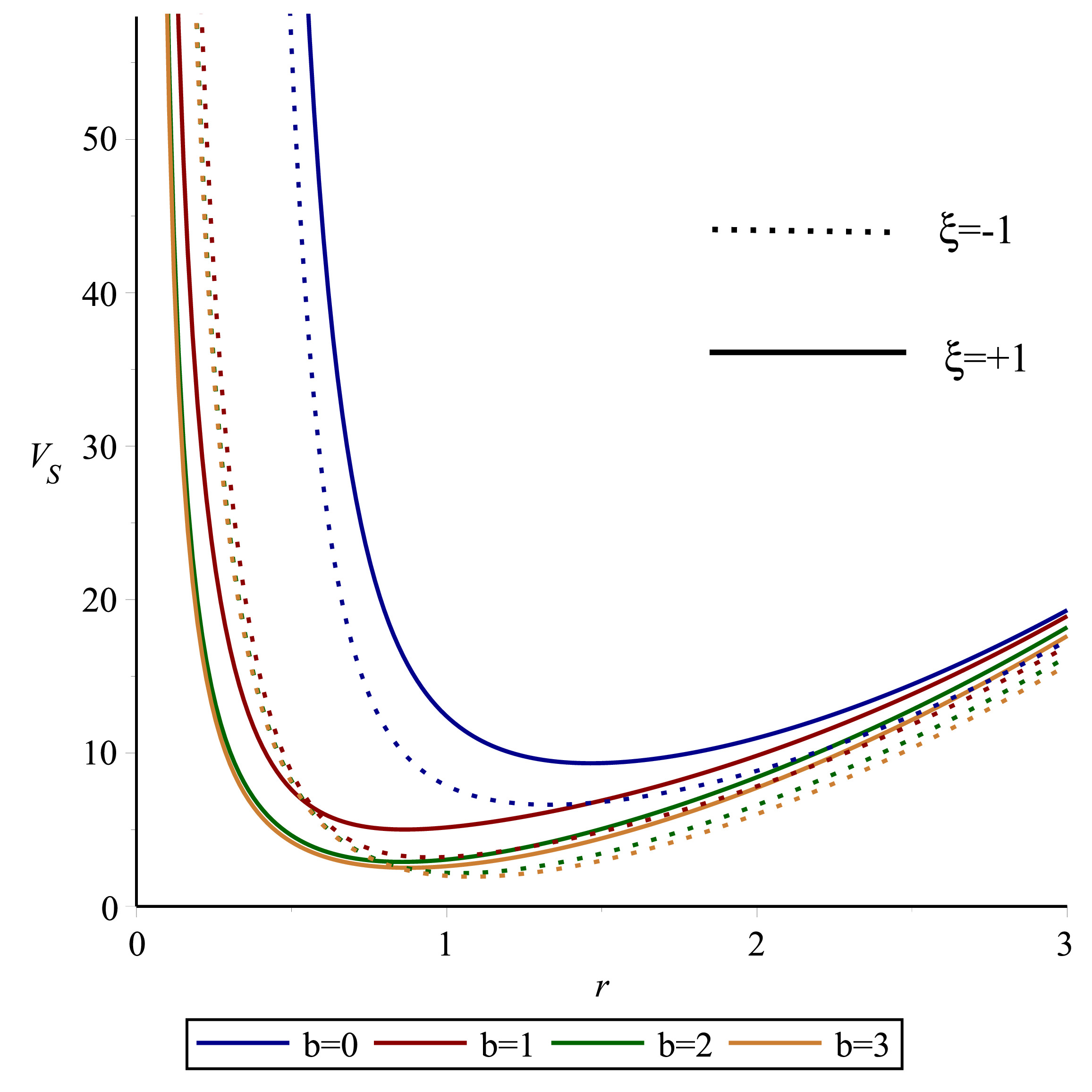}\quad\quad
  \includegraphics[width=.45\textwidth]{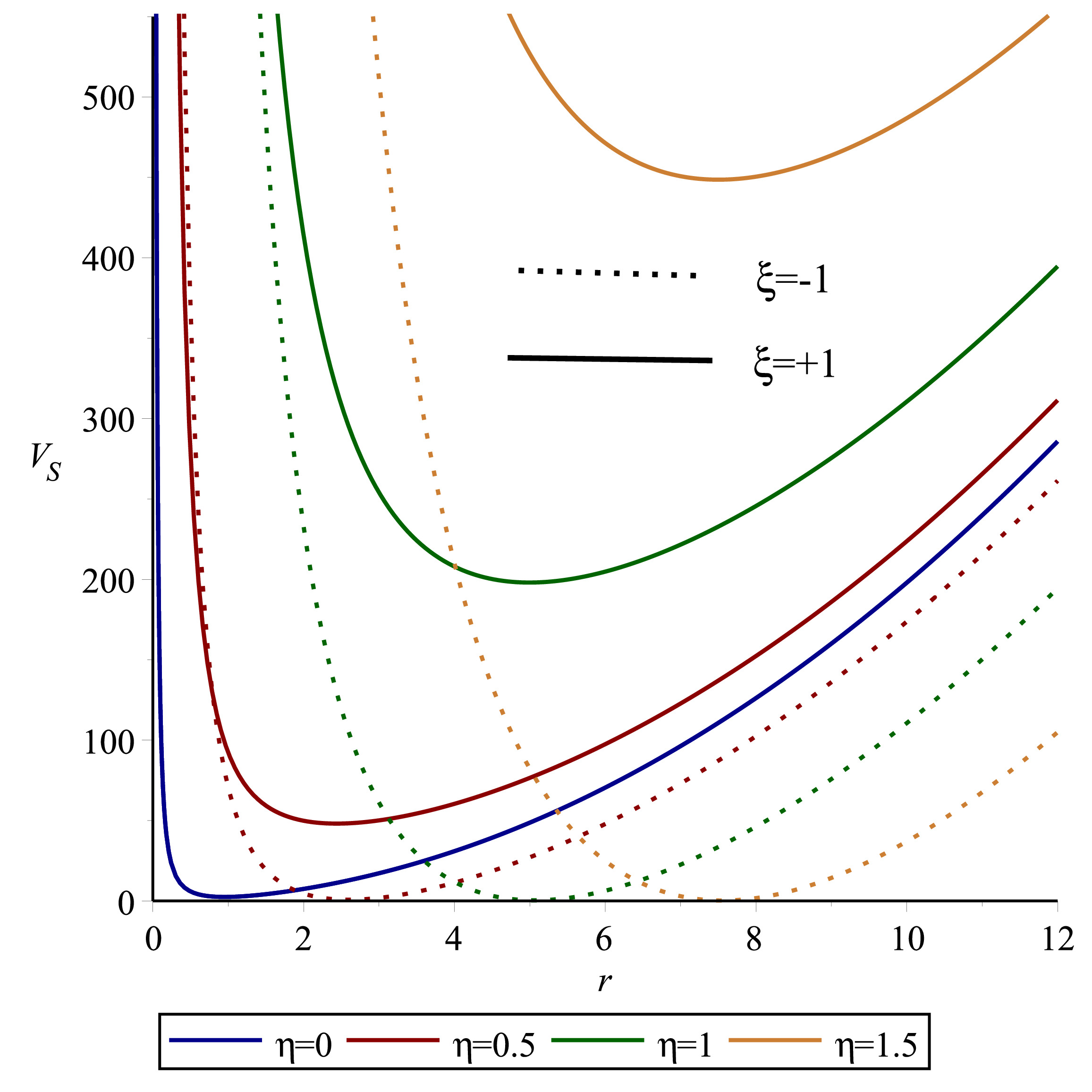}
\caption{RW-potential affected by Bardeen parameter $b$ (left) and the energy scale of the symmetry breaking $\eta$ (right), for the non-zero spin scalar field.} \label{FS3}
\end{figure*}

Fig. \eqref{FS3} demonstrates the RW-potential $\mathcal{V}_S$ behavior by altering parameters $b$ (left) and $\eta$ (right) for $\xi=\pm 1$. With a standard global monopole, $\xi=1$, the RW-potential is given by {\small
\begin{eqnarray}
    \mathcal{V}_S=\left(1-8\pi\eta^2 -\frac{2Mr^2}{(r^2+\mathrm{b}^2)^{3/2}}-\frac{\Lambda}{3} r^2\right)\Bigg[\frac{\ell(\ell+1)}{r^2}+\frac{S(S-1)}{r^2}\left(-8\pi\eta^2-\frac{2Mr^2}{(r^2+\mathrm{b}^2)^{3/2}}-\frac{\Lambda}{3} r^2\right)
    +\frac{(1-S)}{r}\left(\frac{2M(r^2-2\mathrm{b}^2)}{(r^2+\mathrm{b}^2)^{5/2}}-\frac{2\Lambda}{3}\right) \Bigg].\label{cc8}
\end{eqnarray}
}
We see from the above expression (\ref{cc8}) that the presence of an ordinary global monopole in the chosen BH metric decreases the RW-potential of perturbations in comparison to the Bardeen-like AdS BH spacetime.

In the presence of a phantom global monopole, $\xi=-1$, the RW-potential reads {\small
\begin{eqnarray}
    \mathcal{V}_S=\left(1+8\pi\eta^2-\frac{2Mr^2}{(r^2+\mathrm{b}^2)^{3/2}}-\frac{\Lambda}{3} r^2\right)\,\Bigg[\frac{\ell(\ell+1)}{r^2}+\frac{S(S-1)}{r^2}\left(8\pi\eta^2-\frac{2Mr^2}{(r^2+\mathrm{b}^2)^{3/2}}-\frac{\Lambda}{3} r^2\right)
    +\frac{(1-S)}{r}\left(\frac{2M(r^2-2\mathrm{b}^2)}{(r^2+\mathrm{b}^2)^{5/2}}-\frac{2\Lambda}{3}\right) \Bigg].\label{cc9}
\end{eqnarray}
}
The expression (\ref{cc9}) indicates that incorporating a phantom global monopole in the BH metric enhances the RW-potential of perturbations relative to a Bardeen-like AdS BH spacetime.

Below, we present results of the RW-potential for various spin fields.

\subsection{\large \bf Spin-zero scalar perturbations}

In this part, we analyze the RW potential for spin-zero scalar fields. Therefore, for $S=0$, the RW-potential with ordinary global monopole
\begin{eqnarray}
    \mathcal{V}_0=\left(1-8\,\pi\, \eta^2-\frac{2\,M\, r^2}{(r^2+\mathrm{b}^2)^{3/2}}-\frac{\Lambda}{3} r^2\right)\,\Bigg[\frac{\ell\,(\ell+1)}{r^2}+\frac{1}{r}\,\left(\frac{2\,M\, (r^2-2\,\mathrm{b}^2)}{(r^2+\mathrm{b}^2)^{5/2}}-\frac{2\,\Lambda}{3}\right)\Bigg].\label{dd1}
\end{eqnarray}
In the presence of a phantom global monopole, this potential becomes
\begin{eqnarray}
    \mathcal{V}_0=\left(1+8\,\pi\, \eta^2-\frac{2\,M\, r^2}{(r^2+\mathrm{b}^2)^{3/2}}-\frac{\Lambda}{3} r^2\right)\,\Bigg[\frac{\ell\,(\ell+1)}{r^2}+\frac{1}{r}\,\left(\frac{2\,M\, (r^2-2\,\mathrm{b}^2)}{(r^2+\mathrm{b}^2)^{5/2}}-\frac{2\,\Lambda}{3}\right)\Bigg].\label{dd1b}
\end{eqnarray}

\subsection{\large \bf Spin-one vector field perturbations}

In this part, we analyze the RW-potential for spin-1 tensor fields. Therefore, for $S=1$, we obtain the RW-potential with ordinary global monopole as follows
\begin{eqnarray}
    \mathcal{V}_1=\left(1-8\,\pi\, \eta^2-\frac{2\,M\, r^2}{(r^2+\mathrm{b}^2)^{3/2}}-\frac{\Lambda}{3} r^2\right)\,\frac{\ell\,(\ell+1)}{r^2}.\label{ee1}
\end{eqnarray}
With a phantom global monopole, this potential is given by
\begin{eqnarray}
    \mathcal{V}_1=\left(1+8\,\pi\, \eta^2-\frac{2\,M\, r^2}{(r^2+\mathrm{b}^2)^{3/2}}-\frac{\Lambda}{3} r^2\right)\,\frac{\ell\,(\ell+1)}{r^2}.\label{ee1b}
\end{eqnarray}

\subsection{\large \bf Spin-two tensor field perturbations}

Herein, we scrutinize the RW-potential for spin-2 tensor fields. Consequently, for $S=2$, the RW potential associated with a standard global monopole from Eq. (\ref{cc8}) simplifies to:($\mathrm{a}=1-\eta^2$):
\begin{eqnarray}
    \mathcal{V}_2=\left(\mathrm{a}-\frac{2\,M\, r^2}{(r^2+\mathrm{b}^2)^{3/2}}-\frac{\Lambda}{3} r^2\right)\,\Bigg[\frac{\ell\,(\ell+1)}{r^2}
    +\frac{2}{r^2}\,\left(\mathrm{a}-\frac{2\,M\, r^2}{(r^2+\mathrm{b}^2)^{3/2}}-\frac{\Lambda}{3} r^2-1\right)
    -\frac{1}{r}\left(\frac{2\,M\, (r^2-2\,\mathrm{b}^2)}{(r^2+\mathrm{b}^2)^{5/2}}-\frac{2\,\Lambda}{3}\right) \Bigg].\label{ff1}
\end{eqnarray}
While in the presence of a phantom global monopole, the RW-potential from Eq. (\ref{cc8}) simplifies to ($\mathrm{a'}=1+\eta^2$):
\begin{eqnarray}
    \mathcal{V}_2=\left(\mathrm{a}'-\frac{2\,M\, r^2}{(r^2+\mathrm{b}^2)^{3/2}}-\frac{\Lambda}{3} r^2\right)\,\Bigg[\frac{\ell\,(\ell+1)}{r^2}
    +\frac{2}{r^2}\,\left(\mathrm{a}'-\frac{2\,M\, r^2}{(r^2+\mathrm{b}^2)^{3/2}}-\frac{\Lambda}{3} r^2-1\right)
    -\frac{1}{r}\left(\frac{2\,M\, (r^2-2\,\mathrm{b}^2)}{(r^2+\mathrm{b}^2)^{5/2}}-\frac{2\,\Lambda}{3}\right) \Bigg].\label{ff1b}
\end{eqnarray}

\subsection{\large \bf QNM Frequencies}

In this section, the QNM frequencies of a scalar field (Eq. (\ref{cc7}) with $S=0$) are calculated. We shall use the higher-order WKB approximation \cite{Konoplya,konoplya2} with a higher value of $\ell$. The reason for selecting a slightly higher value for the multipole moment $\ell$ is that the inaccuracy associated with the WKB approach drops dramatically as $\ell$ increases. Tables \ref{taba1} and \ref{taba2} summarize the results.  

\begin{center}
\begin{tabular}{|c|c|c|}
 \hline 
   \multicolumn{3}{|c|}{ ($\mathrm{b}=0.5$, $\ell=4$, $n=1$, $M=1$, $\Lambda=-0.002$)}
\\ \hline 
 $\eta$ & $\xi =1$ & $\xi=-1$  \\ \hline
$0$ & $0.906834-0.281432i$ & $0.906834-0.28143i$ \\ 
$0.1$ & $0.892539-0.276199i$ & $0.921246-0.2867i$ \\ 
$0.2$ & $0.850345-0.260727i$ & $0.965188-0.30271i$ \\ 
$0.3$ & $0.782285-0.235738i$ & $1.04084-0.330018i$ \\ 
$0.4$ & $0.691645-0.202576i$ & $1.15204-0.369275i$ \\ 
$0.5$ & $0.582902-0.163342i$ & $1.30475-0.420611i$ \\ 
$0.6$ & $0.461882-0.121012i$ & $1.50788-0.481904i$
\\
\hline
\end{tabular}
\captionof{table}{The QNM of the Ads-Bardeen  BH with Phantom Global Monopoles  for  specific choices of $\eta$.} \label{taba1}
\end{center}

\begin{center}
\begin{tabular}{|c|c|c|}
 \hline 
    \multicolumn{3}{|c|}{ ( $\eta=0.4$, $\ell=4$, $n=1$, $M=1$, $\Lambda=-0.002$)}
\\ \hline 
 $\mathrm{b}$ & $\xi =1$ & $\xi=-1$
\\ \hline 
$0.1$ & $0.568336-0.166706i$ & $1.19993-0.457376i$ \\ 
$0.2$ & $0.570535-0.166238i$ & $1.21396-0.453723i$ \\ 
$0.3$ & $0.573351-0.165618i$ & $1.23272-0.448325i$ \\ 
$0.4$ & $0.577423-0.164682i$ & $1.26168-0.438678i$ \\ 
$0.5$ & $0.582902-0.163342i$ & $1.30475-0.420611i$ \\ 
$0.6$ & $0.590013-0.161456i$ & $1.36855-0.379946i$\\
 \hline
\end{tabular}
\captionof{table}{The QNM of the Ads-Bardeen BH with Phantom Global Monopoles  for  specific choices of $\mathrm{b}$.} \label{taba2}
\end{center}

\begin{figure}[ht!]
    \centering
    \includegraphics[width=0.9\linewidth]{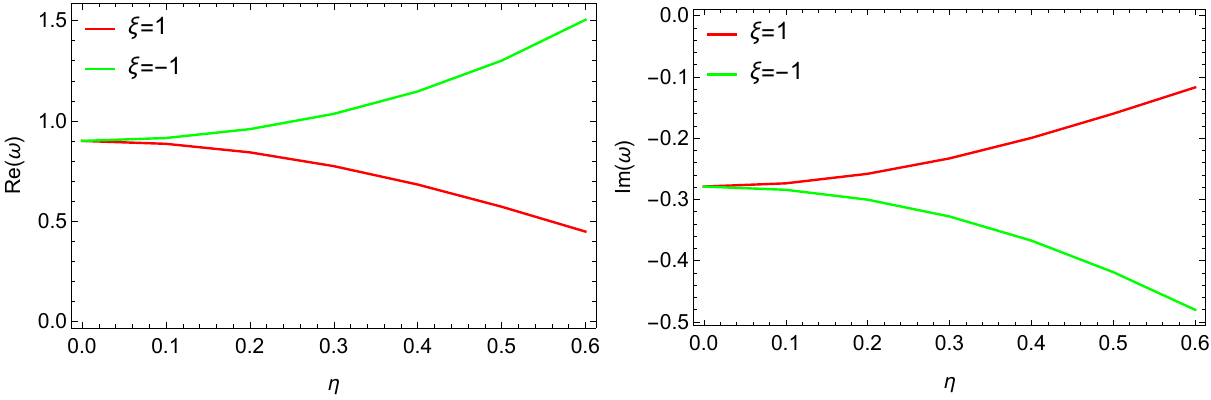}   
    \caption{\textcolor{black}{Plot of QNM frequencies versus $\eta$,  real part (left) and imaginary (right). Here, $\ell=2$}}
    \label{figA1}
\end{figure}

\begin{figure}[ht!]
    \centering
    \includegraphics[width=0.9\linewidth]{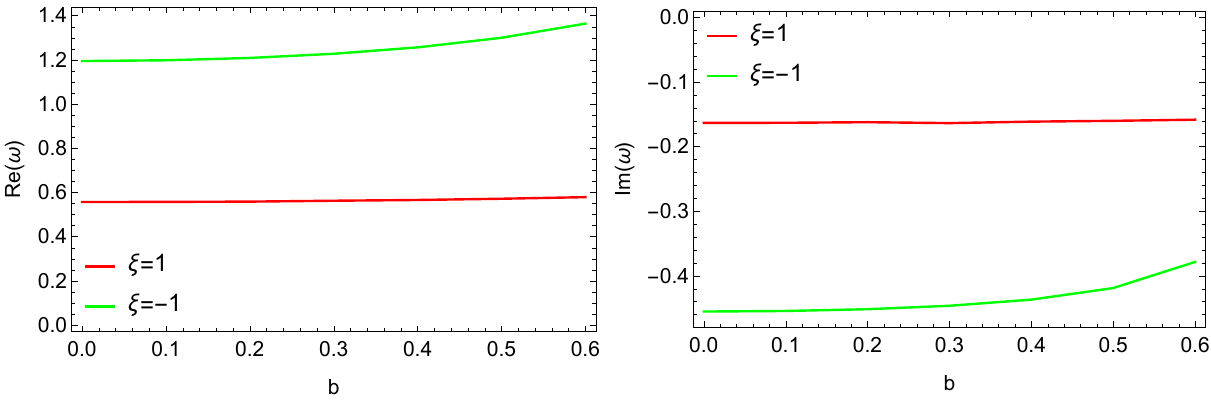}
    \caption{\textcolor{black}{Plot of QNM frequencies versus $\mathrm{b}$, real part  (left) and imaginary part (right). Here, $\ell=2$}}
    \label{figA2}
\end{figure}

Tables \ref{taba1} and \ref{taba2} demonstrate that the imaginary component is negative in all QNM frequency modes. This suggests that the Ads-Bardeen BH with phantom global monopoles returns to steady state after disturbance. To demonstrate how the monopoly parameter $\xi$ and Bardeen parameter $b$ affect the QNM spectrum, Figures \ref{figA1} and \ref{figA2} plot the real and imaginary QNM frequencies.\\ The real component of the QNM frequency, $Re(w)$, determines the oscillation frequency of perturbations. $Im(w)$, the imaginary part of the QNM frequency, determines the damping of perturbations and thus the rate of oscillation decay. In the presence of an ordinary global monopole $\xi=1$ and a fixed Bardeen parameter, $Re(w)$ reduces monotonically as $\eta$ increases. However, when $\eta$ increases, the imaginary part becomes less negative. This indicates that when the energy scale of symmetry breaking $\eta$ increases, the frequency of the oscillation decreases, and the decay of the perturbations slows. For fixed $\eta$, $Re(w)$ increases monotonically as $b$ increases. However, when $b$ increases, the imaginary part becomes less negative.\\In the presence of a phantom global monopole $\xi=-1$ and a fixed Bardeen parameter, $Re(w)$ increases monotonically as $\eta$ increases. However, when $\eta$ increases, the imaginary part becomes more negative. We conclude that the frequency of the oscillation increases and the perturbations decay rapidly when the energy scale of the symmetry breaking $\eta$ increases. For fixed $\eta$, $Re(w)$ increases monotonically as $b$ increases. However, when $b$ increases, the imaginary part becomes less negative. 

\section{\large Transmission and reflection probability}\label{sec4}

This section aims to investigate the transmission and reflection coefficients of the Bardeen-like AdS BH with a PGM. In this study, we apply a semianalytic method that provides a general rigorous bound in the transmission and reflection probabilities for one-dimensional potential scattering \cite{Visser:1998ke,Konoplya:2011,Kanzi:2021jrl,Boonserm:2008dk}. The general bound equation for the transmission and reflection coefficients are given by
\begin{equation}\label{kk1}
    T(\omega)\ge\sec h^2\left(\int_{-\infty}^{+\infty}\wp\, dr_*\right),   
\end{equation}
and 
\begin{equation}\label{kk2}
    R(\omega)\le\tan h^2\left(\int_{-\infty}^{+\infty}\wp\, dr_*\right),   
\end{equation}

here $\wp$ defined as
\begin{equation}\label{kk3}
    \wp=\frac{\sqrt{(H^{\prime})^2+(\omega^2-V_\text{eff}-H^2)^2}}{2H}.
\end{equation}
where $H$ is a positive function with two conditions: 1)$H(r_{*})>0$ and 2) $H(+\infty)=H(-\infty)=\omega$. 

Using either the geodesic effective potential or the RW-potential obtained in Eq. \eqref{bb4} and Eq. \eqref{cc7} respectively, has been crashed due to the undefined terms in both transmission and reflection probability Eq. \eqref{kk1} and Eq. \eqref{kk2}. In this regard, without losing generality, we can set $H^2=\omega^2-V_\text{eff}$ in Eq. \eqref{kk3}. Thus, Eq. \eqref{kk1} and \eqref{kk2} are written by 
\begin{equation}\label{kk4}
    T(\omega)\ge\sec h^2\left(\frac{1}{2}\int_{-\infty}^{+\infty}\Bigg|\frac{H^{\prime}}{H}\Bigg|\, dr_*\right).  
\end{equation}
And 
\begin{equation}\label{kk5}
    R(\omega)\le\tan h^2\left(\frac{1}{2}\int_{-\infty}^{+\infty}\Bigg|\frac{H^{\prime}}{H}\Bigg|\, dr_*\right),  
\end{equation}

The result in parentheses after integration is $\ln{(\frac{H_\text{peak}}{H})}$, thus the transmission and reflection probabilities are given by
\begin{equation}\label{kk6}
    T(\omega)\ge\frac{4\,\omega^2\,(\omega^2-V_\text{peak})}{(2\,\omega^2-V_\text{peak})^2},\quad\quad\quad V_\text{peak}=V^\text{peak}_\text{eff}.
\end{equation}
And 
\begin{equation}\label{kk7}
    R(\omega)\le\frac{V_\text{peak}^2}{(2\,\omega^2-V_\text{peak})^2}.
\end{equation}

\begin{figure}[ht!]
  \includegraphics[width=.4\textwidth]{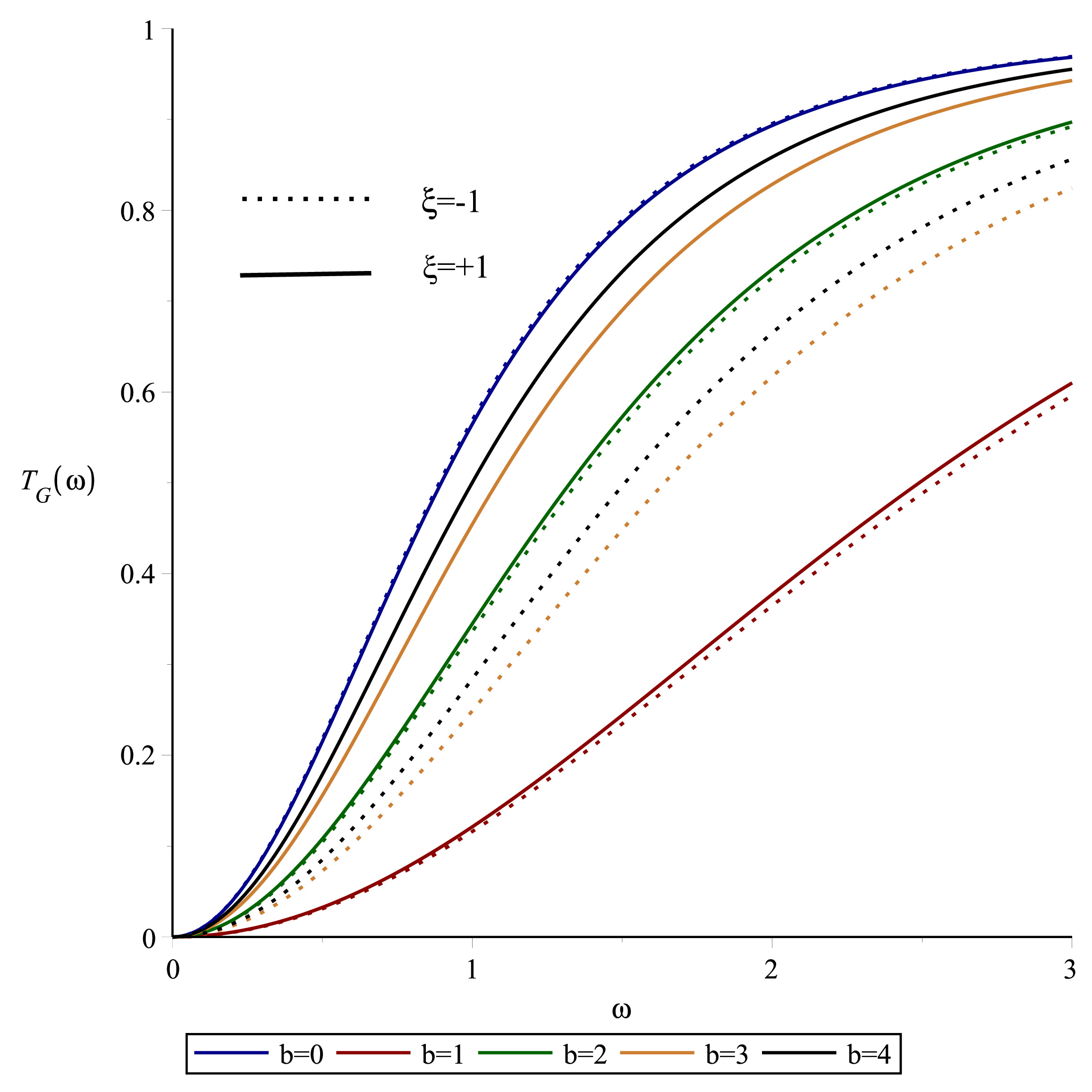}\quad\quad\quad
  \includegraphics[width=.4\textwidth]{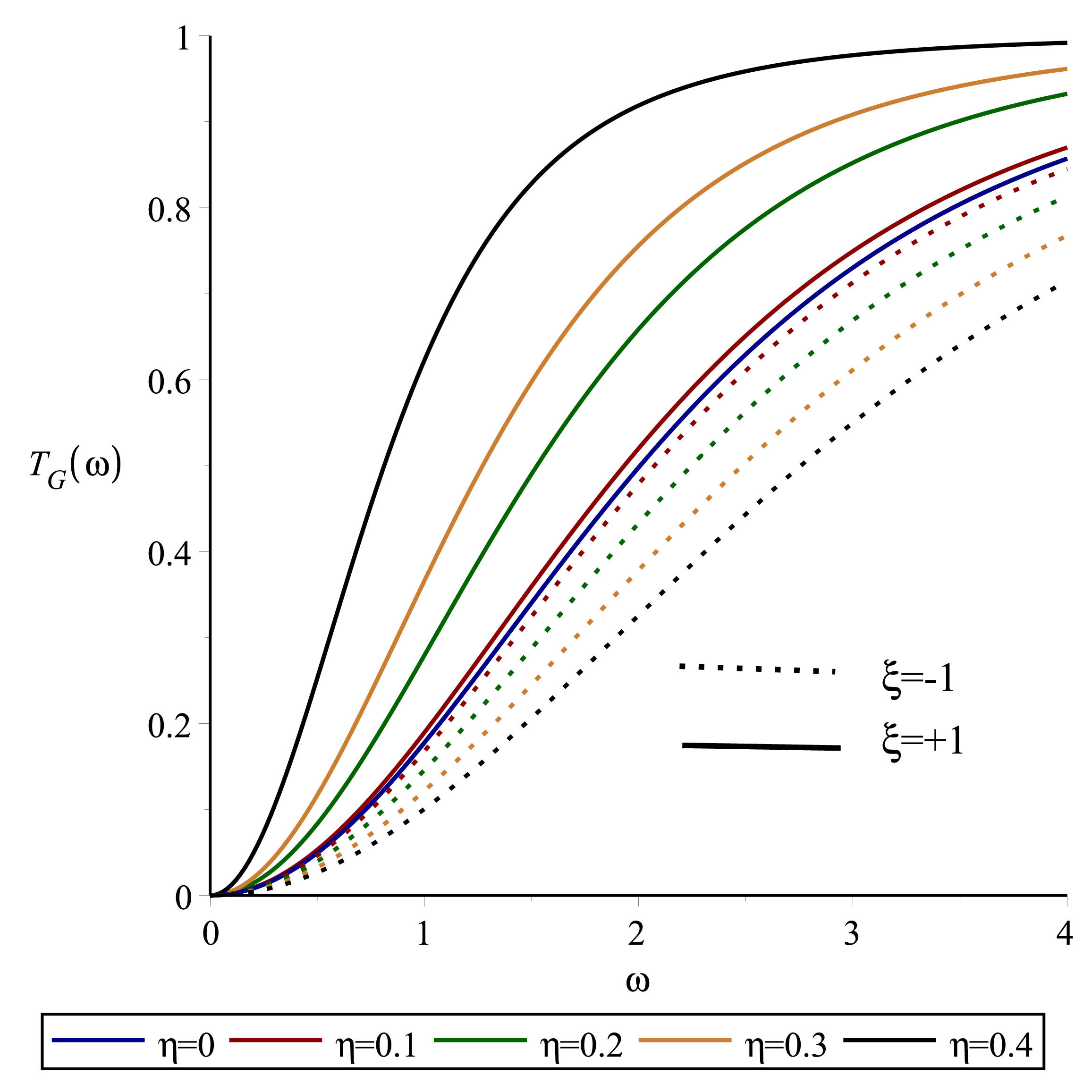}
\caption{Null Geodesics greybody factor under impressions of the Bardeen parameter $b$ (left) and the energy scale $\eta$ (right) for both $\xi=+1$ (solid lines) and $\xi=-1$ (dotted lines). The rest of the arguments are defined as follows: $M=1, \Lambda=3$ and $\ell=2$.} \label{F3}
\end{figure}

\begin{figure}[ht!]
  \includegraphics[width=.4\textwidth]{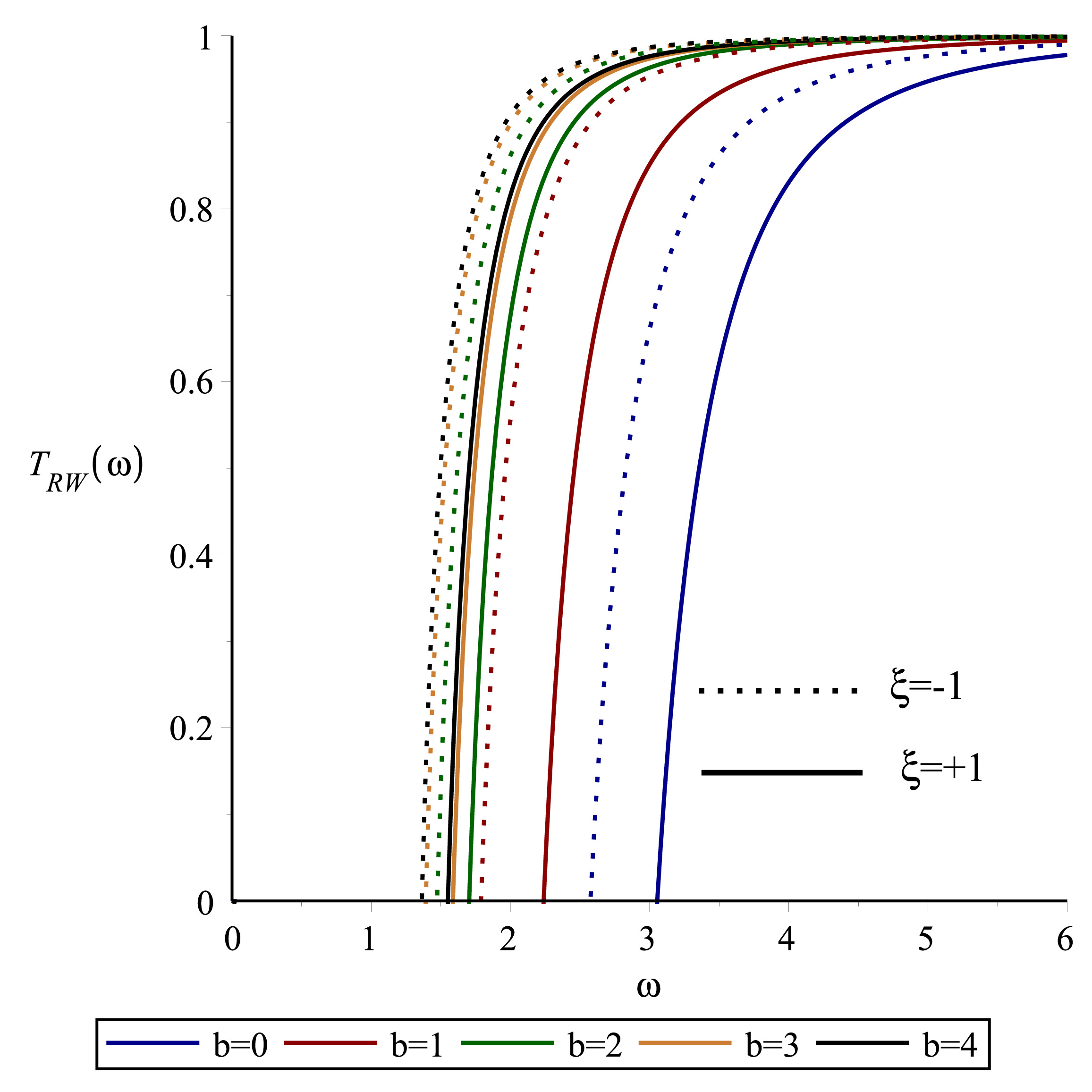}\quad\quad\quad
  \includegraphics[width=.4\textwidth]{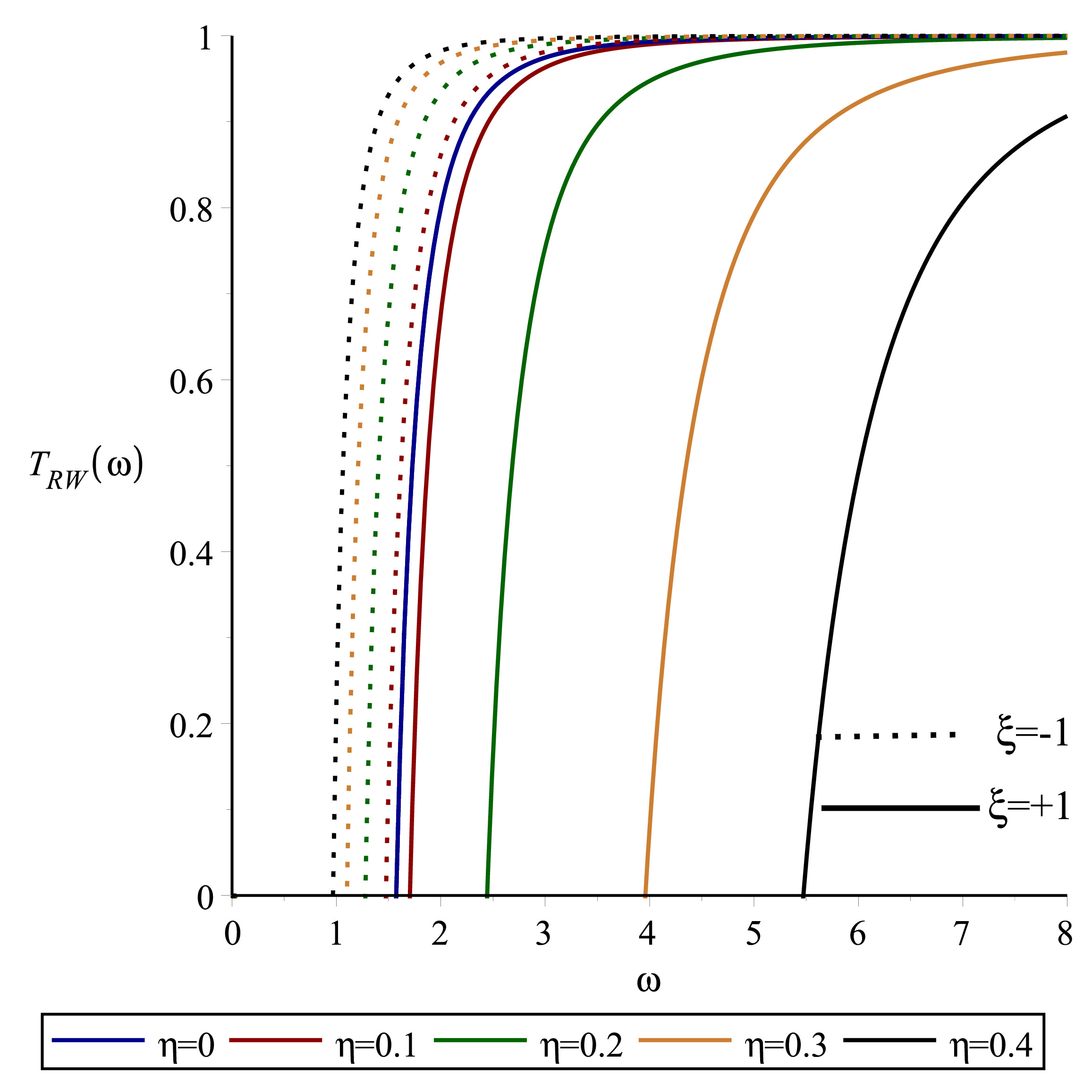}
\caption{RW greybody factor under impression of the Bardeen parameter $b$ (left) and the energy scale $\eta$ (right) for both $\xi=+1$ (solid lines) and $\xi=-1$ (dotted lines).  The rest of the arguments are defined as follows: $M=1, \Lambda=3,s=2$ and $\ell=1$.} \label{F4}
\end{figure}

\begin{figure}[ht!]
  \includegraphics[width=.4\textwidth]{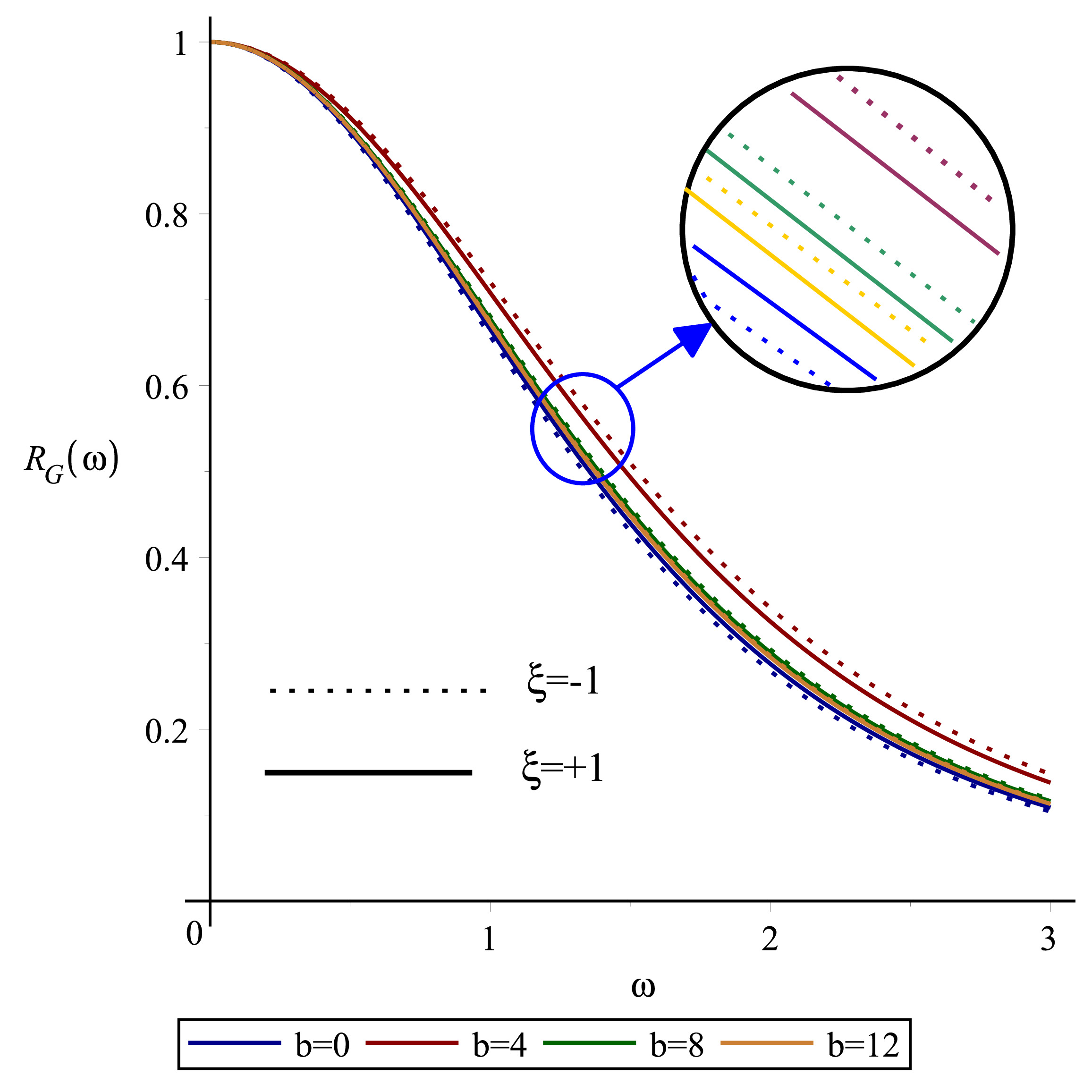}\quad\quad\quad
  \includegraphics[width=.4\textwidth]{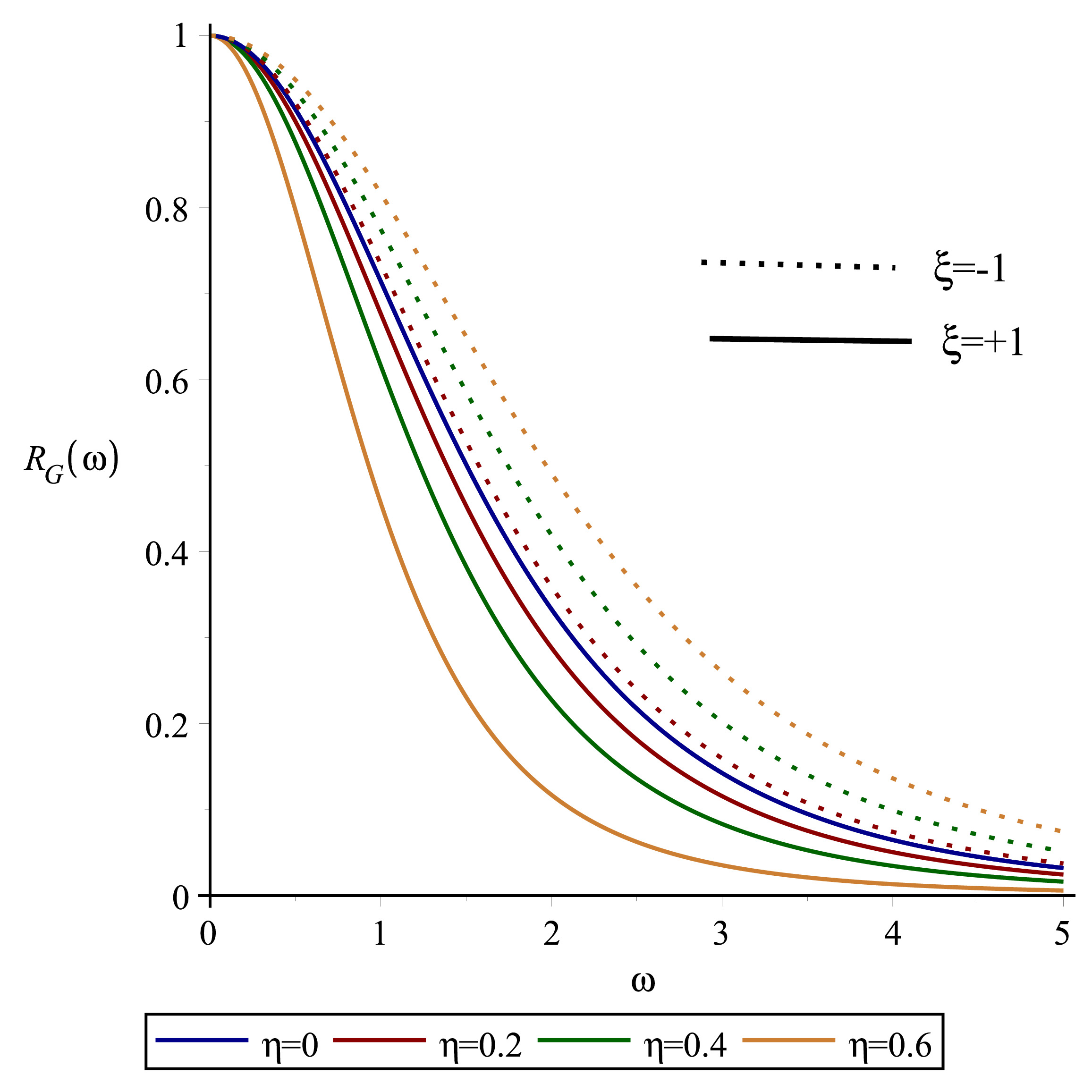}
\caption{Null Geodesics reflected factor under impressions of the Bardeen parameter $b$ (left) and the energy scale $\eta$ (right) for both $\xi=+1$ (solid lines) and $\xi=-1$ (dotted lines). The rest of the arguments are defined as follows: $M=1, \Lambda=3$ and $\ell=2$.} \label{F5}
\end{figure}

\begin{figure}[ht!]
  \includegraphics[width=.4\textwidth]{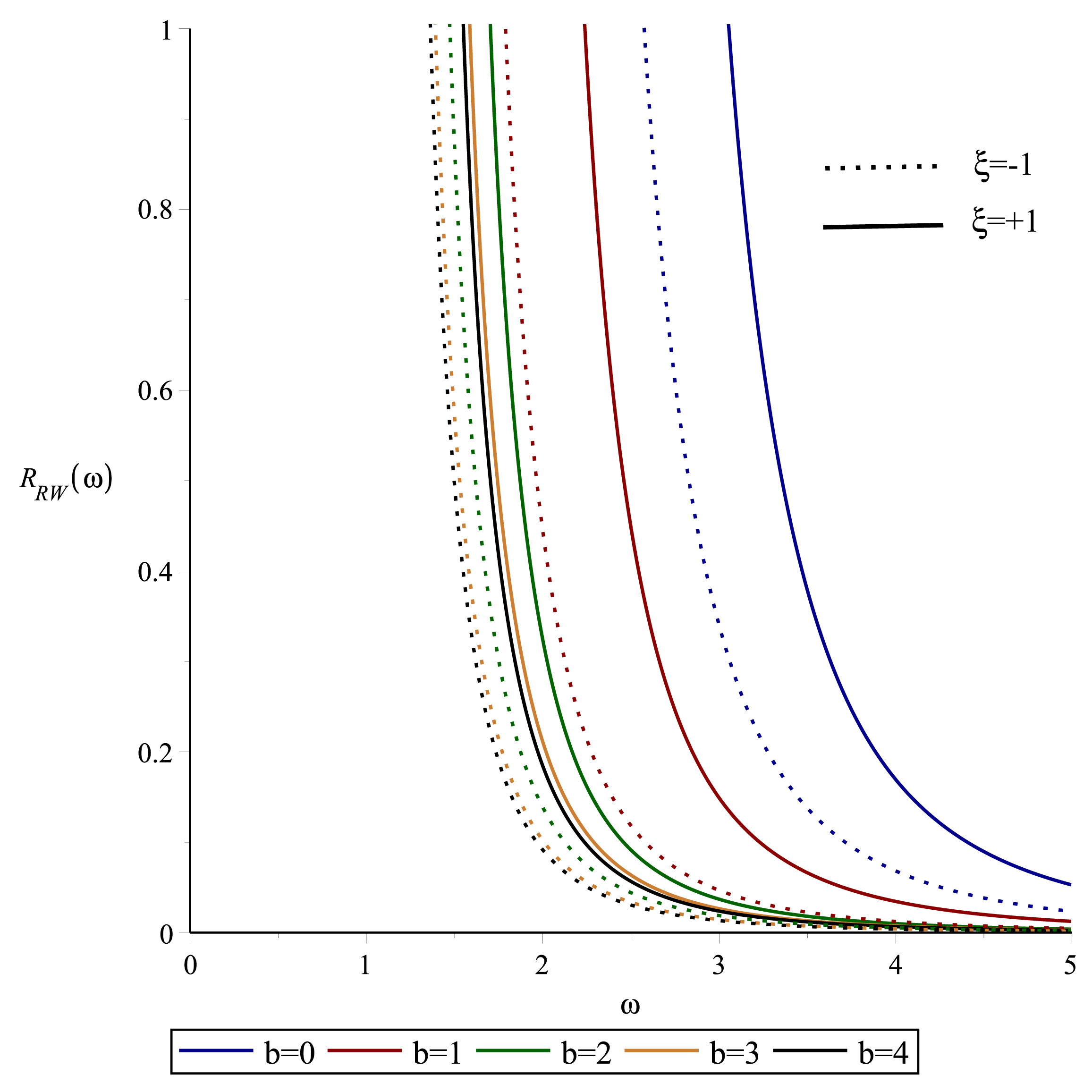}\quad\quad\quad
  \includegraphics[width=.4\textwidth]{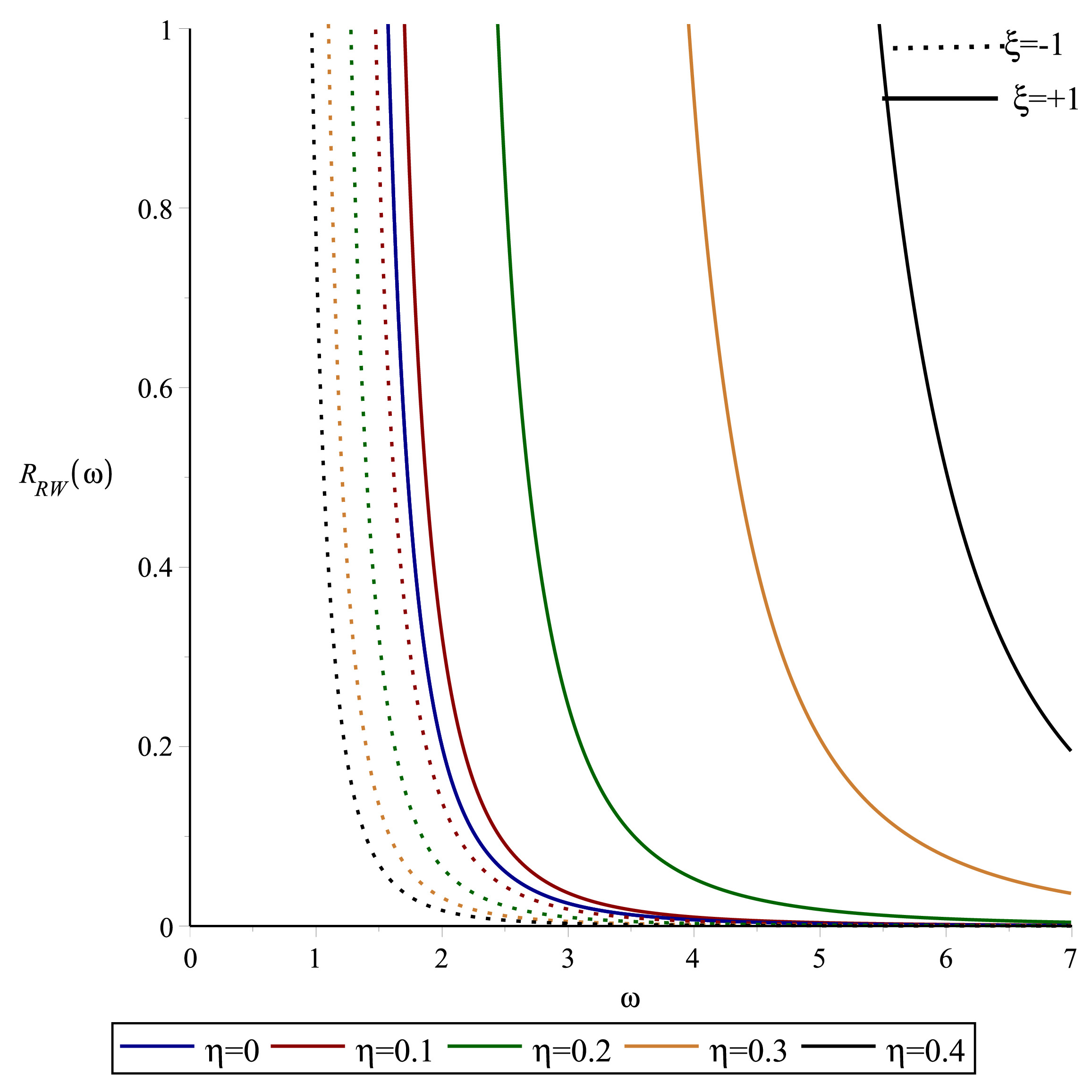}
\caption{RW reflected factor under impression of the Bardeen parameter $b$ (left) and the energy scale $\eta$ (right) for both $\xi=+1$ (solid lines) and $\xi=-1$ (dotted lines).  The rest of the arguments are defined as follows: $M=1, \Lambda=3,s=2$ and $\ell=1$.} \label{F6}
\end{figure}

In order to derive $V_\text{peak}$ in Eq. \eqref{kk6} and Eq. \eqref{kk7} let's first determine the $r_\text{peak}$ by taking the derivative from the geodesics effective potential in Eq. \eqref{bb4} and the RW-potential in Eq. \eqref{cc7}. The results of the direct transmission probability (greybody factor) are depicted in Fig. \eqref{F3} and Fig. \eqref{F4}, and the reflection probability result is illustrated in Fig. \eqref{F5} and Fig. \eqref{F6}. Fig. \eqref{F3} shows the null $(\epsilon=0)$ geodesic transmission probability for the Bardeen parameter $b$ (left) and the energy scale (right) for both $\xi=+1$ (solid lines) and $\xi=-1$ (dotted lines). As shown, the transmission probability for $b=0$ and $b=1$ has the highest and lowest values, respectively. Seemingly, the energy scale $\eta$ has a positive impression on the transmission probability in both $\xi=\pm1$.

The RW transmission probability Fig. \eqref{F4} that is differentiated by $T_{RW}(\omega)$ in addition to a small growth towards one, it is also shifted to the left by increasing the Bardeen parameter $b$ in which this procedure is inverse by increasing the energy scale $\eta$ in the case of $\xi=+1$. It would be significant to mention that in the case of $\xi=+1$ for $\eta>0.1$ the geodesic transmission probability experiences a complex form of $V_{peak}$, in which the real terms are used. 

The behavior of the reflection probability under the influence of the Berdeen parameter $b$, and the energy scales $\eta$ is consistent with the transmission probability in both geodesics and the RW cases, the parameter that causes the increase in the transmission probability should reduce the reflection probability, which is clearly shown in Fig. \eqref{F5}. The probability of geodesic reflection under the impression of the energy scale $\eta$ also experienced the complex $V_{peak}$ for $\eta\ge 0.2$ only for $\xi=+1$, which could be predicted by evaluating the behavior of the effective geodesic potential drawn in Fig. \eqref{FS2} (right). Furthermore, the RW reflection probability depicted in Fig. \eqref{F6} is in harmony with their corresponding transmission probability.

\section{\large \bf Concluding remarks}\label{sec5}

In this study, we thoroughly investigated the dynamics of test particles and perturbations in a Bardeen-like AdS BH spacetime with a PGM. Our analysis considered the impacts of the symmetry-breaking energy scale $\eta$, the Bardeen parameter $b$, and the cosmological constant $\Lambda$ on the geodesic motion, QNMs, and greybody factors. By combining numerical and analytical techniques, we presented a comprehensive exploration of the physical characteristics of this exotic BH solution. 

We began by examining the geodesic motions of null and time-like particles in the presence of the PGM. The effective potential, $V_{\text{eff}}$, governing these motions was derived and analyzed in detail. Our results, illustrated in Figures~\ref{FS2} and \ref{FS3}, showed that the inclusion of the PGM significantly alters the trajectories of particles. Specifically, the energy scale $\eta$ and the Bardeen parameter $b$ introduced distinct deviations in the geodesics, influencing the stability and escape conditions of particles. These findings underscore the importance of such parameters in shaping the spacetime geometry and particle dynamics.

In the context of axial and scalar perturbations, we derived the spin-dependent RW-potential, $\mathcal{V}_S$, which plays a pivotal role in determining the QNM spectrum. Our analysis demonstrated how the RW potential varies with $\eta$, $b$, and $\xi$ (the monopole parameter), providing insights into the stability of the BH under perturbations. We computed the QNMs using the sixth-order WKB approximation, and our results, summarized in Tables~\ref{taba1} and \ref{taba2}, revealed the complex interplay between these parameters and the oscillatory and damping behavior of the perturbations. For example, we observed that increasing $\eta$ decreases the oscillation frequency but slows down the damping rate for $\xi = +1$. Conversely, for $\xi = -1$, higher values of $\eta$ increased the oscillation frequency while accelerating the decay of perturbations.

One of the novel aspects of our work involved the computation of greybody factors through the transmission and reflection probabilities for scalar and axial perturbations. Using semi-analytical methods, we derived bounds on these probabilities and presented their dependence on $\eta$, $b$, and $\xi$ in Figures~\ref{F3} through \ref{F6}. The results indicate that the Bardeen parameter $b$ and the symmetry-breaking scale $\eta$ strongly influence the transmission and reflection probabilities, with higher values of $b$ leading to reduced transmission and increased reflection. These findings have significant implications for understanding the observational signatures of such BHs, particularly in the context of Hawking radiation and wave scattering phenomena.

Future work will focus on extending this framework to include rotating Bardeen-like AdS BHs with PGMs, allowing us to explore the interplay between angular momentum and the exotic parameters. Additionally, we aim to investigate the impact of higher-order corrections in nonlinear electrodynamics and extend the analysis to other modified theories of gravity. The exploration of greybody factors and QNMs in these contexts will further enhance our understanding of the observational imprints of such BHs. We also plan to analyze the impact of quantum gravity corrections \cite{Feng:2021, Pourhassan:2022auo, Pourhassan:2019} on the thermodynamic stability and phase transitions of these systems, potentially providing a more profound understanding of the link between quantum mechanics and GR.

\section*{Acknowledgments}

F.A. acknowledges the Inter University Centre for Astronomy and Astrophysics (IUCAA), Pune, India for granting visiting associateship. \.{I}.~S. thanks EMU, T\"{U}B\.{I}TAK, ANKOS, and SCOAP3 for funding and acknowledges the networking support from COST Actions CA22113, CA21106, and CA23130.

\appendix
{\color{black}
\section{Action and Justification for the Bardeen-like AdS BH with PGM} \label{appendix}

The corresponding action that gives rise to this solution is a combination of Einstein gravity, NLED, and a triplet phantom scalar field theory \cite{CLAR,MBAV,SC}
\begin{equation}
    S = \int d^4x \sqrt{-g} \left[ \frac{1}{16\pi} (R - 2\Lambda) - \frac{1}{4\pi} \mathcal{L}_{\text{NLED}} - \frac{1}{2}\, \xi\, g^{\mu\nu} (\partial_\mu \psi^a)(\partial_\nu \psi^a)- \frac{\lambda}{4} (\psi^a \psi^a - \eta^2)^2 \right].
\end{equation}
To verify that the given metric satisfies the Einstein field equations derived from the proposed action, one can compute the corresponding Einstein field equations:
\begin{equation}
    G_{\mu\nu} + \Lambda g_{\mu\nu} = 8\pi (T^{\text{NLED}}_{\mu\nu} + T^{\text{PGM}}_{\mu\nu}),
\end{equation}
where the stress-energy tensor consists of contributions from NLED and the phantom global monopole field.

\subsection*{1. Computation of the Ricci Tensor and Einstein Tensor}
For the given spherically symmetric metric,
\begin{equation}
    ds^2=-\mathcal{F}(r) dt^2+\mathcal{F}(r)^{-1} dr^2+r^2(d\theta^2+\sin^2\theta d\phi^2),
\end{equation}
the nonzero components of the Ricci tensor are computed as:
\begin{equation}
    R_{tt} = \frac{\mathcal{F}''(r)}{2} - \frac{\mathcal{F}'(r)}{2r} + \frac{\mathcal{F}(r) -1}{r^2},
\end{equation}
\begin{equation}
    R_{rr} = -\frac{\mathcal{F}''(r)}{2} - \frac{\mathcal{F}'(r)}{2r} + \frac{1-\mathcal{F}(r)}{r^2},
\end{equation}
\begin{equation}
    R_{\theta\theta} = 1 - \mathcal{F}(r) + r \mathcal{F}'(r),
\end{equation}
\begin{equation}
    R_{\phi\phi} = \sin^2 \theta R_{\theta\theta}.
\end{equation}

The Einstein tensor components are then obtained as:
\begin{equation}
    G_{tt} = \frac{2(1 - \mathcal{F}(r))}{r^2} + \frac{\mathcal{F}'(r)}{r},
\end{equation}
\begin{equation}
    G_{rr} = -\frac{2(1 - \mathcal{F}(r))}{r^2} - \frac{\mathcal{F}'(r)}{r},
\end{equation}
\begin{equation}
    G_{\theta\theta} = \frac{r \mathcal{F}'(r) + \mathcal{F}(r) - 1}{r^2},
\end{equation}
\begin{equation}
    G_{\phi\phi} = \sin^2 \theta G_{\theta\theta}.
\end{equation}

\subsection*{2. Stress-Energy Tensor Contributions}
The stress-energy tensor for NLED is given by\cite{CLAR}:
\begin{equation}
    T^{\text{NLED}}_{\mu\nu} = \left( \mathcal{L}_{\text{NLED}} - 2F \frac{\partial \mathcal{L}_{\text{NLED}}}{\partial F} \right) g_{\mu\nu} + 4 \frac{\partial \mathcal{L}_{\text{NLED}}}{\partial F} F_{\mu\lambda} F_{\nu}^{\lambda},
\end{equation}
where the NLED Lagrangian responsible for generating the Bardeen-like term is:
\begin{equation}
    \mathcal{L}_{\text{NLED}} = \frac{3}{16\pi} \frac{b^2}{(r^2+b^2)^{5/2}} F,
\end{equation}
with $F = F_{\mu\nu}F^{\mu\nu}$ being the Maxwell invariant.

The energy-momentum tensor for the phantom global monopole is:
\begin{equation}    T_{\mu\nu}^{\text{PGM}} =-\xi\,g^{\mu\nu}\,(\partial_{\mu} \psi^a)(\partial_{\nu} \psi^a)- g_{\mu\nu} \mathcal{L}_{\text{PGM}},
\end{equation}
where the monopole contribution is given by the Lagrangian \cite{MBAV,SC}:
\begin{equation}
    \mathcal{L}_{\text{PGM}} =\frac{\xi}{2}\,g^{\mu\nu}\,(\partial_{\mu} \psi^a)(\partial_{\nu} \psi^a)- \frac{\lambda}{4} (\psi^a \psi^a - \eta^2)^2.
\end{equation}
Thus, action \eqref{izaction}, when varied with respect to the metric, yields the Einstein field equations with the given BH solution \eqref{bb1}.
}

\end{document}